%% file: P1804_E.tex
\documentclass[usegraphicx,useAMS,usenatbib]{mn2e}

\usepackage{aas_macros}
\usepackage[svgnames]{xcolor}
\usepackage{acronym}
\usepackage{xspace}
\usepackage{amsmath,amssymb,amsfonts,amscd}
\usepackage{hyperref}

\usepackage{booktabs} 
\usepackage{siunitx} 
\usepackage{pgfplotstable} 

\newcommand{\modl}[1]{model \texttt{#1}}
\newcommand{\modls}[1]{models \texttt{#1}}
\newcommand{\Modl}[1]{Model \texttt{#1}}
\newcommand{\Modls}[1]{Models \texttt{#1}}
\newcommand{\modelname}[1]{\texttt{#1}}

\newcommand{\nusp}{neutrinosphere\xspace}
\newcommand{\nusps}{neutrinospheres\xspace}

\newcommand{\banel}[1]{\textit{#1}}
\newcommand{\Erot}{\mathcal{T}}
\newcommand{\Emag}{\mathcal{B}}
\newcommand{\pnss}{\textsc{pns}}
\newcommand{\EE}{\textsc{e}}

\newcommand{\Erotpns}{{\cal T}^{\mathrm{\pnss}}}

\newcommand{\Mmax}{M_{\rm bry}^{\rm max}}

\newcommand{\tauAlf}{\tau_{\textrm{Alf}}}
\newcommand{\tauAnu}{\tau_{\textrm{A}\nu}}
\newcommand{\tauAnubl}{\tau_{\textrm{A}\nu; \textrm{bl}}}
\newcommand{\tauadv}{\tau_{\textrm{adv}}}
\newcommand{\tauhtg}{\tau_{\textrm{heat}}}
\newcommand{\tauent}{\tau_{\textrm{ent}}}
\newcommand{\tauhtgbl}{\tau_{\textrm{heat;bl}}}
\newcommand{\tautauavg}{\langle \tauadv / \tauhtg \rangle}
\newcommand{\tautaumavg}{\langle \tauadv / \tauAlf \rangle}
\newcommand{\tautau}{\tauadv / \tauhtg}

\newcommand{\tautauth}{\tauadv (\theta) / \tauhtg (\theta)}

\newcommand{\tautaumth}{\tauadv (\theta) / \tauAlf (\theta)}
\newcommand{\tautauentth}{\tauadv (\theta) / \tauent (\theta)}
\newcommand{\tautauAnu}{\tauadv (\theta) / \tauAnu (\theta)}
\newcommand{\tautauAnubl}{\tauadv (\theta) / \tauAnubl (\theta)}
\newcommand{\Egain}{E_{\textrm{gain}}}
\newcommand{\dotMsh}{F^{\textsc{m}}_{\mathrm{gain}}}

\newcommand{\Alfven}{Alfv{\'e}n\xspace}
\newcommand{\Alfvenic}{Alfv{\'e}nic\xspace}

\newcommand{\pns}{proto neutron star\xspace}

\newcommand{\ie}{i.e.\xspace}

\newcommand{\eg}{e.g.\xspace}
\newcommand{\viz}{viz.\xspace}
\newcommand{\wrt}{w.r.t.\xspace}

\newcommand{\msol}{M_{\odot}}
\newcommand{\msun}{$M_{\odot}$}

\newcommand{\Msol}{M_{\odot}}

\newcommand{\zehn}[1]{10^{#1}}
\newcommand{\zehnh}[2]{{#1} \times 10^{#2}}

\newcommand{\figref}[1]{Fig.\,\ref{#1}}
\newcommand{\tabref}[1]{Tab.\,\ref{#1}}
\newcommand{\secref}[1]{Sect.\,\ref{#1}}

\newcommand{\third}{$3^{\mathrm{rd}}$}
\newcommand{\nth}[1]{${#1}^{\mathrm{th}}$}

\newcommand{\ms}{\textrm{ms}}

\newcommand{\km}{\textrm{km}}
\newcommand{\cm}{\textrm{cm}}
\newcommand{\cms}{\textrm{cm s}^{-1}}
\newcommand{\erg}{\textrm{erg}}

\newcommand{\sek}{\textrm{s}}

\newcommand{\gccm}{\textrm{g\,cm}^{-3}}

\newcommand{\grad}{^{\circ}}

\newcommand{\MeV}{\textrm{MeV}}

\newcommand{\Alnum}{\mathsf{A}}

\newcommand{\jcop}{JCP}

\let\oldhref\href
\renewcommand{\href}[2]{\oldhref{#1}{\hbox{#2}}}

\begin{document}

\title
{Magnetorotational core collapse of possible GRB progenitors.
  I. Explosion mechanisms.}

\author[Obergaulinger \& Aloy]{
  M.~Obergaulinger$^1$,  M.\'A.~Aloy$^2$
  \\
  $^1$ Institut f{\"u}r Kernphysik, Theoriezentrum, S2|11
  Schlo{\ss}gartenstr.~2, 64289 Darmstadt, Germany
  \\
  $^2$ Departament d'Astronomia i Astrof{\'i}sica, Universitat de
  Val{\`e}ncia,  Edifici Jeroni Munyoz, C/
  Dr.~Moliner, 50, E-46100 Burjassot (Val{\`e}ncia), Spain 
}

\maketitle

\begin{abstract}
  We investigate the explosion of stars with zero-age main-sequence
  masses between 20 and 35 \msun and varying degrees of
  rotation and magnetic fields including ones commonly considered
  progenitors of gamma-ray bursts (GRBs). The simulations, combining
  special relativistic magnetohydrodynamics, a general relativistic
  approximate gravitational potential, and two-moment neutrino
  transport, demonstrate the viability of different scenarios for the
  post-bounce evolution. Having formed a highly massive proto-neutron
  star (PNS), several models launch successful explosions, either by
  the standard supernova mechanism based on neutrino heating and
  hydrodynamic instabilities or by magnetorotational processes. It is,
  however, quite common for the PNS to collapse to a black hole (BH)
  within a few seconds. Others might produce proto-magnetar-driven
  explosions. We explore several ways to describe the different
  explosion mechanisms. The competition between the timescales for
  advection of gas through the gain layer and heating by neutrinos
  provides an approximate explanation for models with insignificant
  magnetic fields. The fidelity of this explosion criterion in the
  case of rapid rotation can be improved by accounting for the strong
  deviations from spherical symmetry and mixing between pole and
  equator. We furthermore study an alternative description including
  the ram pressure of the gas falling through the shock. Magnetically
  driven explosions tend to arise from a strongly magnetised region
  around the polar axis. In these cases, the onset of the explosion
  corresponds to the equality between the advection timescale and the
  timescale for the propagation of \Alfven waves through the gain
  layer.
\end{abstract}

\begin{keywords}
  Supernovae: general - gamma-ray bursts: general
\end{keywords}

\section{Introduction}
\label{Sek:Intro}

The collapse of the core of a massive star of more than about 8 solar
masses at the end of its hydrostatic evolution is the starting point
for a complex sequence of events with many possible outcomes
\citep[for reviews, see,
\eg,][]{Janka_et_al__2016__AnnualReviewofNuclearandParticleScience__PhysicsofCore-CollapseSupernovaeinThreeDimensionsASneakPreview,Mueller__2016__pasa__TheStatusofMulti-DimensionalCore-CollapseSupernovaModels}. 
The post-collapse evolution depends strongly on factors such as
the mass and metallicity of the progenitor, its rotation and magnetic
field, and potentially on the individual realization of stochastic
processes such as hydrodynamic instabilities.  During the subsequent
period of up to several seconds, neutrinos streaming out of the \pns (PNS)
transfer energy to the gas behind the stalled shock wave and, together
with hydrodynamic instabilities, and possibly other effects such as
rotation and magnetic fields, favour shock revival.  For a successful
core-collapse supernova (CCSN) explosion to occur, they have to unbind
at least parts of the matter surrounding the PNS and overcome the
continuous infall of matter at high velocities towards the PNS.  The
properties of the explosion such as mass, composition, velocity,
energy, and geometry of the ejecta depend strongly on the progenitor
and on the relative importance of different processes contributing to
shock revival.  This view is corroborated by observations of a wide
variety of CCSNe ranging from low-luminosity events to very energetic
ones characterized by high expansion velocities and pronounced global
asymmetries \citep[see, \eg][]{Woosley_Bloom__2006__araa__The_Supernova_Gamma-Ray_Burst_Connection,Cano_et_al__2017__AdvancesinAstronomy__TheObserversGuidetotheGamma-RayBurstSupernovaConnection,Moriya_et_al__2018__ssr__SuperluminousSupernovae}.

We are interested in the core collapse of stars in a regime where all
of the aforementioned evolutionary paths overlap.  The goal is to
study the physics of explosions of rapidly rotating stars at different
metallicities, with different magnetic fields and initial masses and
to explore the conditions for the development of long GRB engines.
More specifically, we focus on the following issues:
\begin{enumerate}
\item Do stars commonly considered likely long GRB progenitors explode
  in CCSNe?
\item If so, what mechanism produces the shock revival and what are
  the characteristics of the explosion?
\end{enumerate}

We aim to model stars likely to produce high-mass PNSs while at the
same time close to the threshold between successful explosion and
failed shock revival.  Based on the aforementioned considerations, we
select several stars of $20$ and $35 \, \msol$
\citep{Woosley_Heger_Weaver__2002__ReviewsofModernPhysics__The_evolution_and_explosion_of_massive_stars,Woosley_Heger__2006__apj__TheProgenitorStarsofGamma-RayBursts,Woosley_Heger__2007__physrep__Nucleosynthesisandremnantsinmassivestarsofsolarmetallicity}
that, while differing in their metallicity, and, in case of stars that
were evolved including the effects of rotation and magnetic fields,
also the mass loss rates, are all close to the explosion threshold.

Our progenitors differ by their mass, structure, rotation and magnetic
field from those studied most extensively in supernova theory.
Consequently, we expect deviations from the standard scenario for
explosions driven by neutrino heating that is aided by hydrodynamic
instabilities \citep[for a review, see,
\eg][]{Janka__2012__ARNPS__ExplosionMechanismsofCore-CollapseSupernovae}.
In particular, rotation and possibly magnetic fields are likely to
contribute to the explosion mechanism.  Rapid rotation 
may lead to global asymmetries of
the shock wave, and of the neutrino emission, which
translate into the formation of bipolar outflows
\citep{Obergaulinger_Aloy__2017__mnras__Protomagnetarandblackholeformationinhigh-massstars}.
These effects are most pronounced when rotation is combined with a
strong magnetic field that can tap into the rotational energy, as has
been demonstrated in multi-dimensional simulations with varying
degrees of approximations regarding the microphysics \citep[see,
\eg][]{Bisnovatyi-Kogan_Popov_Samokhin__1976__APSS__MHD_SN,Mueller_Hillebrandt__1979__AA__MHD_SN,Symbalisty__1984__ApJ_MHD_SN,Akiyama_etal__2003__ApJ__MRI_SN,Kotake_etal__2004__Apj__SN-magrot-neutrino-emission,Thompson_Quataert_Burrows__2004__ApJ__Vis_Rot_SN,Moiseenko_et_al__2006__mnras__A_MR_CC_model_with_jets,Obergaulinger_Aloy_Mueller__2006__AA__MR_collapse,Obergaulinger_et_al__2006__AA__MR_collapse_TOV,
  Dessart_et_al__2007__apj__MagneticallyDrivenExplosionsofRapidlyRotatingWhiteDwarfsFollowingAccretion-InducedCollapse,Burrows_etal__2007__ApJ__MHD-SN,Winteler_et_al__2012__apjl__MagnetorotationallyDrivenSupernovaeastheOriginofEarlyGalaxyr-processElements,Sawai_et_al__2013__apjl__GlobalSimulationsofMagnetorotationalInstabilityintheCollapsedCoreofaMassiveStar,Mosta_et_al__2014__apjl__MagnetorotationalCore-collapseSupernovaeinThreeDimensions,Moesta_et_al__2015__nat__Alarge-scaledynamoandmagnetoturbulenceinrapidlyrotatingcore-collapsesupernovae}.
The dynamic relevance of
the magnetic field depends crucially on the ratio of the
magnetic energy to the kinetic energy, which in most, though not all,
typical pre-collapse cores is expected to be rather small.  Hence,
processes that amplify the seed field such as flux-freezing
compression, winding by the differential rotation, or dynamos driven
by the MRI or hydrodynamic instabilities are important ingredients to
the overall picture
\citep{Akiyama_etal__2003__ApJ__MRI_SN,Obergaulinger_etal__2009__AA__Semi-global_MRI_CCSN,Masada_et_al__2012__apj__LocalSimulationsoftheMagnetorotationalInstabilityinCore-collapseSupernovae,Moesta_et_al__2015__nat__Alarge-scaledynamoandmagnetoturbulenceinrapidlyrotatingcore-collapsesupernovae,Guilet_Mueller__2015__mnras__Numericalsimulationsofthemagnetorotationalinstabilityinprotoneutronstars-I.Influenceofbuoyancy,Masada_et_al__2015__apjl__MagnetohydrodynamicTurbulencePoweredbyMagnetorotationalInstabilityinNascentProtoneutronStars,Rembiasz_et_al__2016__JournalofPhysicsConferenceSeries__TerminationoftheMRIviaparasiticinstabilitiesincore-collapsesupernovae:influenceofnumericalmethods,Rembiasz_et_al__2016__mnras__Onthemaximummagneticfieldamplificationbythemagnetorotationalinstabilityincore-collapsesupernovae,Sawai_Yamada__2016__apj__TheEvolutionandImpactsofMagnetorotationalInstabilityinMagnetizedCore-collapseSupernovae,Guilet_et_al__2017__mnras__Magnetorotationalinstabilityinneutronstarmergers:impactofneutrinos}.
Explosions partially modified or predominantly driven by these
processes have
been invoked to explain several very energetic and aspherical events
associated to hypernovae
\citep{Wheeler_Meier_Wilson__2002__ApJ__MHD_SN,Maeda_Nomoto__2003__apj__Bipolar_SN_Nucleosynthesis_and_Implications_for_Abundances_in_EMP_Stars,Dessart_et_al__2008__apjl__TheProto-NeutronStarPhaseoftheCollapsarModelandtheRoutetoLong-SoftGamma-RayBurstsandHypernovae,Tominaga__2009__apj__Aspherical_Properties_of_Hydrodynamics_and_Nucleosynthesis_in_Jet-Induced_Supernovae,Dessart_et_al__2012__mnras__Superluminoussupernovae:$56$Nipowerversusmagnetarradiation,Mazzali_et_al__2014__mnras__Anupperlimittotheenergyofgamma-rayburstsindicatesthatGRBs/SNearepoweredbymagnetars,Wang_et_al__2015__apj__SuperluminousSupernovaePoweredbyMagnetars:Late-timeLightCurvesandHardEmissionLeakage,Tchekhovskoy_Giannios__2015__mnras__Magneticfluxofprogenitorstarssetsgamma-rayburstluminosityandvariability,Metzger_et_al__2015__mnras__Thediversityoftransientsfrommagnetarbirthincorecollapsesupernovae,Chen_et_al__2016__apj__Magnetar-PoweredSupernovaeinTwoDimensions.I.SuperluminousSupernovae}.

In order to undertake the previously mentioned point (ii), we compare
the evolution across numerical simulations, where we vary the rotation
profiles and the magnetic fields of our pre-collapse cores.
Our simulations are based on a state-of-the-art code combining
high-order methods for solving the hyperbolic terms of the MHD and
transport equations with a post-Newtonian treatment of gravity, as
well as a spectral two-moment neutrino transport including corrections
due to the velocity (Doppler shifts, aberration) and the gravitation
field (gravitational blue/redshift) and the relevant reactions between
neutrinos and matter.  The rather long simulation times we want to
reach (various seconds post-bounce) and the variety of models we need
to explore limit us to consider only two low-resolution
three-dimensional (3D) models, which are prototypes of collapsar- and
PM-forming central engines. These cases present qualitatively the same
behaviour in 3D than in 2D. Encouraged by the similarity of the
results, but aware of the fact that the final answers can only come
from unrestricted 3D models, we explore many other cases employing
axisymmetric models.

This article is organized as follows: the physical model and the
numerical code will be outlined in \secref{Sek:PhysNum}, followed by
an overview of the initial models in \secref{Sek:Init}.  We present
the results of our models in \secref{Sek:Res} and a summary and the
conclusions in \secref{Sek:SumCon}.

\section{Physics and numerics}
\label{Sek:PhysNum}

The simulations were performed using the neutrino-MHD code presented
in
\cite{Just_et_al__2015__mnras__Anewmultidimensionalenergy-dependenttwo-momenttransportcodeforneutrino-hydrodynamics}.
With respect to the description of the algorithms, implementation, and
tests given there \citep[and
in][]{Rembiasz_et_al__2017__apjs__OntheMeasurementsofNumericalViscosityandResistivityinEulerianMHDCodes}
and to the previous application in simulations of magnetized core
collapse
\citep{Obergaulinger_et_al__2014__mnras__Magneticfieldamplificationandmagneticallysupportedexplosionsofcollapsingnon-rotatingstellarcores},
we have made several modifications for the purpose of running
the present set of models. These modifications have already been used
in  closely related publications
\citep[][Paper II hereafter]{Obergaulinger_Aloy__2017__mnras__Protomagnetarandblackholeformationinhigh-massstars,Aloy_Obergaulinger_2019_II},
as well as in the exploration of magneto-rotational collapse of lower
mass and solar metallicity progenitors
\citep{Obergaulinger_et_al__2018__JournalofPhysicsGNuclearPhysics__Corecollapsewithmagneticfieldsandrotation}. Below
we detail all these modifications for the sake of completness.

The MHD system and the hyperbolic part of neutrino transport are
solved in a finite-volume discretization in spherical coordinates,
$r, \theta, \phi$, assuming axisymmetry.  We employ the
constrained-transport method for avoiding a non-zero divergence of the
magnetic field \citep{Londrillo_Del_Zanna__2004__JCP__UpwindCT} and
use high-resolution shock-capturing methods combining high-order
reconstruction \citep[the monotonicity-preserving method of \nth{5}
order (MP5) of][]{Suresh_Huynh__1997__JCP__MP-schemes} and approximate
Riemann solvers \citep[for MHD, HLLC, see][for the neutrinos,
HLL]{Mignone_Bodo__2006__MNRAS__HLLC-RMHD}.  We employ an explicit
\third-order Runge-Kutta time integrator for all terms but the stiff
source terms, which our code integrates implicitly \cite[for the
implementation,
see][]{Just_et_al__2015__mnras__Anewmultidimensionalenergy-dependenttwo-momenttransportcodeforneutrino-hydrodynamics}.

In the following, we will use natural units, where both the speed of
light in vacuum, $c$, and the gravitational constant $G$ are taken to
be $G=c=1$. Furthermore, roman indices $i$, $j$ and $k$ run along the
three spatial dimensions, $1,2,3$, while $m=1,\ldots,N_{\rm spec}$
annotates the species number.

\subsection{Special relativistic MHD}

Here, in contrast to our earlier work based on Newtonian MHD, we solve
the equations of special relativistic MHD, i.e.~the conservation laws
for relativistic mass density, $D$, partial densities of charged
particles (electrons and protons), $Y_e D$, and of a set of chemical
elements, $X_{m} D$, relativistic momentum and energy
density, $\vec S$ and $\tau$, respectively, and magnetic field,
$\vec B$,\footnote{We express all 3-vectors in orthonormal bases,
    so that covariant and contravariant components are
    interchangeable.}
\begin{eqnarray}
  \label{Gl:MHD-rho}
  \partial_{t} D 
  + \vec \nabla \alpha D \vec v 
  & = & 0,
  \\ 
  \label{Gl:MHD-Yrho}
  \partial_{t} Y_e D 
  + \vec \nabla \alpha Y_e D \vec v 
  & = & \alpha Q_{\star}^{Y_e},
  \\ 
  \label{Gl:MHD-Ycmp}
  \partial_{t} X_m D 
  + \vec \nabla \alpha X_m D \vec v 
  & = & R_m,
  \\ 
  \label{Gl:MHD-mom}
  \partial_{t} S^i
  + \nabla_j \alpha \mathcal{T}^{ij}
  & = & \alpha Q^i_{\star} - D \nabla^i \Phi, 
  \\ 
  \label{Gl:MHD-erg}
  \partial_{t} \tau
  + \vec \nabla \alpha \vec F_{\tau}
  & = & \alpha Q^{0}_{\star} - S_i \nabla^i \Phi, 
  \\ 
  \label{Gl:MHD-ind}
  \partial_{t} \vec B  
  + \vec \nabla \times \alpha ( \vec v \times \vec B ) 
  & = & 0,
  \\ 
  \label{Gl:MHD-divb}
  \vec \nabla \cdot \vec B  & =  & 0.
\end{eqnarray}
The operator
$\nabla_i = \frac{1}{\sqrt{\gamma}}\partial_i \sqrt{\gamma}$
($i=1,2,3$) contains the determinant of the spatial metric, $\gamma$,
which does not depend on time.  The fluxes are functions of the
\emph{primitive} variables: velocity, $\vec v$, and Lorentz factor,
$W = (1 - v^2)^{-1/2}$, rest-mass density, $\rho = D / W$, electron
fraction, $Y_e$, and gas pressure, $P$.

The relations between conserved and primitive variables are
\begin{eqnarray}
  \label{Gl:MHD-prim-rho}
  D & = & \rho W,
  \\
  \label{Gl:MHD-prim-S}
  S_i & = & ( \rho h + b^2 ) W^2 - b_i b^0,
  \\
  \label{Gl:MHD-prim-tau}
  \tau & = & ( \rho h + b^2 ) W^2 - ( P + b^2/2 ) - ( b^0)^2 - D,
\end{eqnarray}
where $b^2:=b_\nu b^\nu$ ($\nu=0,\ldots,3$) is the square of the
magnetic field four vector, whose temporal and spatial components are, respectively
\begin{eqnarray}
  \label{Gl:MHD-prim-b0}
  b^0 & = & W B^i v_i,
  \\
  \label{Gl:MHD-prim-bi}
  b^i & = & B^i/W + b^0 v^i.
\end{eqnarray}
For recovering the primitive variables, we use the same techniques
as in \cite{Leismann_etal__2005__AA__RMHD-Jets} or \cite{Cerda-Duran__2008__AA__GRMHD-code}.

Since the relations
inverting Eqs.\,(\ref{Gl:MHD-prim-rho})-(\ref{Gl:MHD-prim-tau}) are
not explicit, the momentum and energy fluxes are given in terms of a
combination of conserved and primitive variables:
\begin{eqnarray}
  \label{Gl:MHD-flux-S}
  \mathcal{T}^{ij} & = & 
  S^j v^i + \delta^{ij} ( P + b^2 / 2 ) - b^j B^i / W, 
  \\
  \label{Gl:MHD-flux-tau}
  F_{\tau}^i & = & 
  \tau v^i + ( P + b^2 / 2) v^i - b^0 B^i / W,
\end{eqnarray}
with $\delta^{ij}$ being the Kroneker delta.  The other quantities
appearing in the MHD equations are the lapse function, $\alpha$, and
the source terms accounting for the exchange of lepton number,
momentum, and energy between matter and neutrinos, $Q_{\star}^{Y_e},
Q^{i}_{\star}$, and $Q^{0}_{\star}$, respectively.  The source terms
denoted with the $\star$ subscript are the integrals over neutrino
energy, summed over all neutrino flavours of the spectral
neutrino-matter interaction terms, which we will discuss below.

In the stars with $M_{\mathrm{ZAMS}} = 35 \, \msol$, the gas pressure
is determined by the equation of state (EOS) of
\cite{Lattimer_Swesty__1991__NuclearPhysicsA__LS-EOS} with a nuclear
incompressibility of $K_3 = 220 \, \MeV$ (LS220 hereafter) for
densities above $\rho_{\rm low} = \zehnh{6}{7} \, \gccm$.  At lower
densities, we use an EOS containing contributions of
electrons and positrons, photons, and baryons and the flashing scheme
of \cite{Rampp_Janka__2002__AA__Vertex}.  The
general way in which
this scheme changes the nuclear composition of the gas is represented
by the set of source terms $R_{m}$.  We employ the same low-density
EOS in the models with $M_{\mathrm{ZAMS}} = 20 \, \msol$, but a
different high-density EOS, \viz SFHo of
\cite{Steiner_et_al__2013__apj__Core-collapseSupernovaEquationsofStateBasedonNeutronStarObservations}.
Because, this EOS is tabulated for a wider range of densities, we
switch between it and the low-density EOS at a lower density of
$\rho_{\rm low} = 6000 \, \gccm$.  We note that the maximum baryonic
mass (in the non-rotating and zero temperature limit) is $\Mmax\simeq
2.45\msol$ for both EoSs (LS220 and SFHo).

We compute the gravitational potential, $\Phi$, according to version
'A' of the post-Newtonian TOV potentials of
\cite{Marek_etal__2006__AA__TOV-potential}.  To ensure consistency
with the gravitational terms in the equations of neutrino transport
(see below), we include the lapse in the spatial derivatives.  Since
we do not model gravity by a general relativistic 3+1 metric, we
define the lapse function based on the classical gravitational
potential, $\Phi$, as $\alpha = \exp ( \Phi / c^2)$.

\subsection{Neutrino transport}

We treat the neutrino transport in the two-moment framework closed by
the maximum-entropy Eddington factor
\citep{Cernohorsky_Bludman__1994__ApJ__MEC-Transport}.  As in
\cite{Obergaulinger_Aloy__2017__mnras__Protomagnetarandblackholeformationinhigh-massstars}
and
\cite{Obergaulinger_et_al__2018__JournalofPhysicsGNuclearPhysics__Corecollapsewithmagneticfieldsandrotation},
we include the effects of gravity in the neutrino-transport equations
in the $\mathcal{O}(v)$-plus formulation of
\cite{Endeve_et_al__2012__ArXive-prints__ConservativeMomentEquationsforNeutrinoRadiationTransportwithLimitedRelativity}:
\begin{eqnarray}
  \label{Gl:neu-erg}
  \partial_{t} E + \partial_{t} v_i F^i
  & + & \vec \nabla \alpha ( \vec F +
  \vec v E )
  \\ 
  \nonumber
  & - & 
  ( \nabla_i \alpha + \dot{v}_i) \left[ \partial_{\epsilon} (\epsilon F^i) - F^i
  \right]
  \\ 
  \nonumber
  & - & 
  \nabla_i ( \alpha v_j )
  \left[ \partial_{\epsilon} ( \epsilon P^{ij}) - P^{ij}\right]
  \\ \nonumber
  & =  &
  \alpha Q_{0},
  \\
  \label{Gl:neu-mom}
  \partial_{t} ( F^i + v_j P^{ij} )
  & + & \nabla_j ( \alpha P^{ij} + v^j F^i) +      \dot{v}^i E
  \\ \nonumber
  & + &
  \alpha F^j \nabla_j v^i
  +      ( E + P^j_j) \nabla^i \alpha
     -   \partial_{\epsilon} (\epsilon P^i_{j}) \dot{v}^j
  \\ \nonumber
  & - & \alpha \partial_{\epsilon} ( \epsilon U^{ki}_j) \nabla_k v^j
       - \partial_{\epsilon} ( \epsilon P^{ij}) \nabla_j \alpha
  \\ \nonumber
  & =  & 
  \alpha Q^i,
\end{eqnarray}
where $P^i_j$ and $U^{ki}_j$ are the second and third moment of the
neutrino distribution function, respectively, and $\dot{\vec v}$ is
the acceleration.  The momentum equation involves the third
moment, for which we use the approximation given in
\cite{Just_et_al__2015__mnras__Anewmultidimensionalenergy-dependenttwo-momenttransportcodeforneutrino-hydrodynamics}.
The terms involving $\alpha$ that appear in addition to the ones
contained in our earlier work are implemented in an analogous fashion
to the formally very similar expressions in the velocity-dependent
terms see \citep[for details,
see][]{Just_et_al__2015__mnras__Anewmultidimensionalenergy-dependenttwo-momenttransportcodeforneutrino-hydrodynamics}.

\subsection{Neutrino-matter interaction}

The most important change with respect to the basic set
of reactions from
\cite{Obergaulinger_et_al__2014__mnras__Magneticfieldamplificationandmagneticallysupportedexplosionsofcollapsingnon-rotatingstellarcores}
consists in the addition of pair processes, electron-positron
annihilation and nucleonic bremsstrahlung, in the implementation of
which we follow
\cite{Pons_Miralles_Ibanez__1998__AAPS__nu_Annihil_cross_section} and
\cite{Hannestad_Raffelt__1998__apj__SupernovaNeutrinoOpacityfromNucleon-NucleonBremsstrahlungandRelatedProcesses},
respectively. Hence, neutrinos of all flavours are produced in our
simulations \cite[in contrast to the ones
of][]{Obergaulinger_et_al__2014__mnras__Magneticfieldamplificationandmagneticallysupportedexplosionsofcollapsingnon-rotatingstellarcores}.
The other reactions included are:
(i) nucleonic absorption, emission, and scattering with the
corrections due to weak magnetism and recoil;
(ii) nuclear absorption, emission, and scattering;
(iii) inelastic scattering off electrons.

\section{Initial models and parameters}
\label{Sek:Init}

\begin{table*}
  \centering
  \begin{tabular}{|l|cccccc|}
    \hline
    star & $M_{\star} \, [\msol]$ & $\rho_{c} \, [\zehn{9} \, \gccm]$ &
    $M_{\mathrm{Fe}} \,    [\msol]$ & $R_{\mathrm{Fe}} \, [\zehn{8}\,
    \cm]$
    & $\Omega_{\mathrm{c}} \, [s^{-1}]$
    & $\Omega_{\mathrm{Fe}} \, [s^{-1}]$
    \\ 
    \hline 
    35OC & 28.1 & 2.4 & 2.0 & 3.0 & 2.0 & 0.1
    \\ 
    35OB & 21.2 & 3.2 & 2.2 & 2.8 & 1.5 & 0.05
    \\ 
    s20 & 14.7 & 5.8 & 1.5 & 1.7 & $-$ & $-$ 
    \\
    z35 & 35.0 & 2.2 & 2.3 & 3.2 & $-$ & $-$ 
    \\ \hline
  \end{tabular}
  \caption{
    Properties of the four stellar models at the onset of collapse:
    the total mass, $M_{\star}$, the central density,
    $\rho_{\mathrm{c}}$, the mass and radius of the iron
    core, $M_{\mathrm{Fe}}$ and $R_{\mathrm{Fe}}$, and the angular
    velocity at the centre and the surface of the iron
    core, $\Omega_{\mathrm{c}}$ and $\Omega_{\mathrm{Fe}}$ ($-$ for
    models evolved without rotation). We note that we use as operative
    definition of the iron core the part of the star where the iron
    fraction is larger than 0.1. This explains the small differences
    in the listed values of $M_{\rm Fe}$ with respect to the ones
    listed in Woosley \& Heger (2006).
  }
  \label{Tab:init}
\end{table*}

As we focus in particular on the effects of rotation and magnetic
fields, it should be pointed out that all progenitor models are the
result of spherically symmetric stellar evolution calculations.  Some
of them (models 35OB/C, see below) include approximate prescriptions
for rotational dynamos and the feedback of the magnetic field on the
stellar structure, whereas others do not incorporate them at all.  In
the former case, we could base the pre-collapse distributions of
angular velocity and magnetic field on the stellar-evolution models.
For the latter class of models, we add different distributions of the
angular velocity and the magnetic field to the spherically symmetric
pre-collapse stellar configurations.  For the sake of comparison, we
employ the same technique for additional simulations of the former
models.  We finally note that some of the simulations have been
presented already
\citep{Obergaulinger_Aloy__2017__mnras__Protomagnetarandblackholeformationinhigh-massstars,Obergaulinger_et_al__2018__JournalofPhysicsGNuclearPhysics__Corecollapsewithmagneticfieldsandrotation},
but some of the models have been evolved further in time.  Here, we
present a more comprehensive investigation of the processes that lead
to a successful explosion or lack thereof.

We use the progenitor model 35OC, which was computed by
\cite{Woosley_Heger__2006__apj__TheProgenitorStarsofGamma-RayBursts}
as a model for a rapidly rotating star of zero-age main-sequence mass
$M_{\mathrm{ZAMS}} = 35 \, \Msol$ including the redistribution of
angular momentum by magnetic fields according to the
theoretical framework of \cite{Spruit__2002__AA__Dynamo}.  We used
another model from the same series of progenitors, model 35OB, which
was constructed assuming a stronger mass loss than model 35OC, leading
to a lower pre-collapse mass and a slower rotating iron core.  The
group of pre-collapse models computed without rotation and magnetic
fields consists of two progenitors with solar metallicity and
$M_{\mathrm{ZAMS}} = 20 \, \msol$ \citep[model
s20][]{Woosley_Heger__2007__physrep__Nucleosynthesisandremnantsinmassivestarsofsolarmetallicity}
and one with zero metallicity and
$M_{\mathrm{ZAMS}} = 35 \, \msol$\citep[model
z35][]{Woosley_Heger_Weaver__2002__ReviewsofModernPhysics__The_evolution_and_explosion_of_massive_stars}.
The properties of these four stars are summarised in
\tabref{Tab:init}.

We map the pre-collapse structure of the core (computed in one spatial
dimension), viz.~the hydrodynamic variables such as density, electron
fraction, temperature, and rotational velocity, onto our simulation
grid.  The magnetic field of the model is the result of MHD
instabilities.  Hence, it is not a global, e.g. dipole field
encompassing the entire star, but rather confined to several shells.
It is given in terms of the absolute values of the toroidal and a
poloidal component as functions of radius only.  Because we lack
detailed information on the orientation of the field vectors as well
as the angular distribution of the field strength, we treat the
magnetic field as an additional parameter of our models.  We simulate
a series of models based on the field as given by the pre-collapse
model 
and an additional series of models in which we replace the field by a
global dipole field and a toroidal component. Assessing the impact of
multipolar magnetic topologies on the post-collapse outcome is beyond
the scope of this paper, but we refer interested readers to the
detailled study of \cite{Bugli_et_al__2019__arXive-prints__Theimpactofnon-dipolarmagneticfieldsincore-collapsesupernovae}.
In the former series of models, we set the $\phi$-component of the
initial field proportional to the toroidal component of the
stellar-evolution model,
\begin{equation}
  \label{Gl:Init-Heger-Bphi}
  b^{\phi} = \beta_0^{\phi} b^{\mathrm{tor}}_{\mathrm{35OC}},
\end{equation}
and compute the $r$-component from its poloidal component 
\begin{equation}
  \label{Gl:Init-Heger-Br}
  b^{r} = \beta_0^{\mathrm{pol}} b^{\mathrm{pol}}_{\mathrm{35OC}} \cos ( n^r
  \theta),
\end{equation}
where $\beta_0^{\mathrm{pol/tor}}$ and $n^r$ are dimensionless
parameters.  The $\theta$-component follows directly from the
solenoidal condition.  In the latter series of models, in which we
lack of magnetic fields in the progenitor or we replace the stellar
evolution field by a large-scale magnetic dipole, we compute the poloidal components from a vector
potential, $\vec A$\citep[e.g.][]{Suwa_etal__2007__pasj__Magnetorotational_Collapse_of_PopIII_Stars}. The
$\phi$-component of $\vec A$ is given in terms of two
parameters, $B_0^{\mathrm{p}}$ and $R_0$,
\begin{eqnarray}
  \label{Gl:init-Suwa-1}
  A^{\phi} & = & B_0^{\mathrm{p}} \frac{R_0^3}{R_0^3 + r^3} r \cos\theta,
\end{eqnarray}
and add a toroidal component proportional to $A^{\phi}$,
\begin{eqnarray}
  \label{Gl:init-Suwa-2}
  b^{\phi} = B_0^{\phi} \frac{R_0^3}{R_0^3 + r^3} r \cos\theta,
\end{eqnarray}

Within each of the two series, we vary the normalization of the
poloidal and toroidal components, and, in the first series, also their
angular distribution.  In order to assess the importance of rotation,
we add a version of the same model with an artificially modified
(either reduced or increased) angular velocity.

All models were simulated on spherical grids. In the case of axially
symmetric models, the mesh consisted of $n_{\theta} = 128$ zones in
$\theta$-direction and $n_r = 400$ radial zones with a width given in
terms of a parameter $(\delta_r)_0 = 600 \, \mathrm{m}$
\begin{equation}
  \label{Gl:radigrid}
  \delta r = \max \left( (\delta r)_0, r \frac{ \pi}{n_{\theta}} \right).
\end{equation}
Compared to other simulations in the literature, the central grid
spacing is relatively coarse, but it is comparable with the values
$(\delta r)_0=500\,$m used by, \eg
\cite{Dimmelmeier_Font_Mueller__2002__AA__GR-Collapse_1}.
We do, however, not see any of the commonly observed artefacts of an
insufficient radial resolution such as a possible expansion of the PNS
surface.  We attribute this fact to our use of high-order
reconstruction schemes, which, as shown by
\cite{Rembiasz_et_al__2017__apjs__OntheMeasurementsofNumericalViscosityandResistivityinEulerianMHDCodes}
greatly reduce the numerical errors \wrt lower-order schemes.  For an
assessment of the behaviour of this code and its dependence on
different assumptions and approximations for the neutrino physics as
well as on the grid resolution, see
\cite{Just_et_al__2018__ArXive-prints__Core-collapsesupernovasimulationsinoneandtwodimensions:comparisonofcodesandapproximations}.

As a result of the setup expressed by Eq.\,\ref{Gl:radigrid}, the grid
width is uniform inside a certain radius and increases outside of this
radius linearly with $r$ in such a manner that the zones have an
aspect ratio very close to unity.  In addition, we apply a coarsening
scheme
close to the origin in order to
increase the time step allowed by the CFL condition and to obtain
effective zones of aspect ratio $\approx 1$, there, too.  To save
computing time, we simulated the collapse of the models (during which
angular resolution is not crucial) in axisymmetry in lower resolution
(the same radial grid, but $n_\theta=32$) and mapped to the standard
grid shortly before bounce.

In energy space, we used $n_{\epsilon} = 10$ energy bins distributed
logarithmically between $\epsilon_{\mathrm{min}} = 3 \, \MeV$ and
$\epsilon_{\mathrm{max}} = 240 \, \MeV$.  This may seem a rather low
resolution, but we made sure by several series of spherically
symmetric tests that our choice does
not affect the results.

The models and their most important properties described in the
following are introduced in \tabref{Tab:models}.

\begin{table}
  \centering
  \begin{tabular}{|l|l|ll|ll|}
    \hline
    name & star & rotation & field & fate & BH
    \\
    \hline
    
    \modelname{35OC-RO} & \modelname{35OC}
    & Or & Or & MR & $+$
    \\
    \modelname{35OC-RO2} & \modelname{35OC}
    & Or & $2\mathrm{p},2\mathrm{t}$ & MR & $\surd$
    \\
    \modelname{35OC-Rp2} & \modelname{35OC}
    & Or & $2\mathrm{p},1\mathrm{t}$ & MR & $\times$
    \\
    \modelname{35OC-Rp3} & \modelname{35OC}
    & Or & $3\mathrm{p},1\mathrm{t}$ & MR & $\times$
    \\
    \modelname{35OC-Rp4} & \modelname{35OC}
    & Or & $4\mathrm{p},1\mathrm{t}$ & MR & $\times$
    \\
    \modelname{35OC-Rw} & \modelname{35OC}
    & Or & $a(10,10)$ & $\nu$-$\Omega$ & ?
    \\
    \modelname{35OC-Rs} & \modelname{35OC}
    & Or & $a(12,12)$ & MR & $\times$
    \\
    \modelname{35OC-Sw} & \modelname{35OC}
    & $\times \frac{1}{4}$ & $a(8,10)$ & $\nu$ & $\surd$
    \\
    \modelname{35OC-RRw} & \modelname{35OC}
    & $\times 0.5$ & Or & $\nu$-$\Omega$ & $\times$
    \\
    \modelname{35OC-RO-TOV} & \modelname{35OC}
    & Or & Or & MR &  $\surd$
    \\
    \hline
    \modelname{35OB-RO} & \modelname{35OB}
    & Or & Or & $\nu$-$\Omega$ & $\surd$
    \\
    \modelname{35OB-RRw} & \modelname{35OB}
    & $\times 2$ & Or$/ 10^{6}$ & $\times$ & $+$
    \\
    \hline
    \modelname{s20-1} & \modelname{s20}
    & R & a(10,11) & $\times$ & $\times$
    \\
    \modelname{s20-2} & \modelname{s20}
    & a(0.1) & a(10,11) & $\nu$ & $\times$
    \\
    \modelname{s20-3} & \modelname{s20}
    & a(1.0) & a(11,11) & MR & $\times$
    \\
    \modelname{s20-2noB} & \modelname{s20}
    & a(0.1) & 0 & $\times$ & $+$
    \\
    \modelname{s20-3noB} & \modelname{s20}
    & a(1.0) & 0 & $\times$ & $+$
    \\
    \hline
    \modelname{z35-Sw} & \modelname{z35.0}
    & a(0.5) & a(8,10) & $\nu$ & $\surd$
    \\
    \modelname{z35-Rw} & \modelname{z35.0}
    & a(1.0) & a(8,10) & $\nu$-$\Omega$ & $+$
    \\
    \hline
  \end{tabular}
  \caption{
    List of our models.
    Each simulation is listed with its name and the progenitor star.
    The third column indicates the type of the
    rotation profile: ``Or'' stands for the original profile taken from the
    stellar evolution calculation, $\times n$ means that we multiplied
    the original angular velocity by a uniform factor $n$, ``R''
    indicates a random velocity field of negligible magnitude, and
    $a(\Omega)$ denotes the artificial $j$-constant rotational profile
    with a central angular velocity of $\Omega$.  
    The fourth column similarly shows the type of magnetic field:
    ``Or'' indicates the magnetic field profile of the original
    stellar evolution model, $x\mathrm{p},y\mathrm{t}$ means that the
    original poloidal and toroidal fields have been multiplied by
    factors $x$ and $y$, respectively, and $\mathrm{a}(x,y)$ stands
    for an artificial field with maximum poloidal and toroidal field
    components of $10^x$ and $10^y$ G, respectively.
    The fifth column, ``fate'', gives a brief indication of the
    evolution of the model: $\nu$ means a standard neutrino-driven
    shock revival, $\nu$-$\Omega$ one strongly affected by rotation, MR a
    magnetorotational explosion, and $\times$ a failed explosion.
    The last column shows the sign $\surd$ if a BH formed during the
    simulation, $+$ if it did not, but we consider its formation
    likely on time scales of seconds after the end of the simulation,
    and $\times$ if no BH was formed and we estimate the final remnant
    to be a NS. The fate of \modl{35OC-Rw} is unclear, hence we annotate
    it with a  question mark.
  }
  \label{Tab:models}
\end{table}

\section{Results}
\label{Sek:Res}

In the following, we will present a general outline of the dynamics of
the models, characterising them according to the aforementioned
evolutionary paths (\secref{sSek:ResOv}).  This presentation will be
followed by a deeper look at several effects that shape the evolution
of the cores (\secref{sSek:ResDe}).  We conclude this section with an
analysis of the results in the light of a few conditions that
have been suggested for shock revival (\secref{sSek:ResEC}).

\subsection{Overview}
\label{sSek:ResOv}

\begin{figure}
  \centering
  \includegraphics[width=0.95\linewidth]{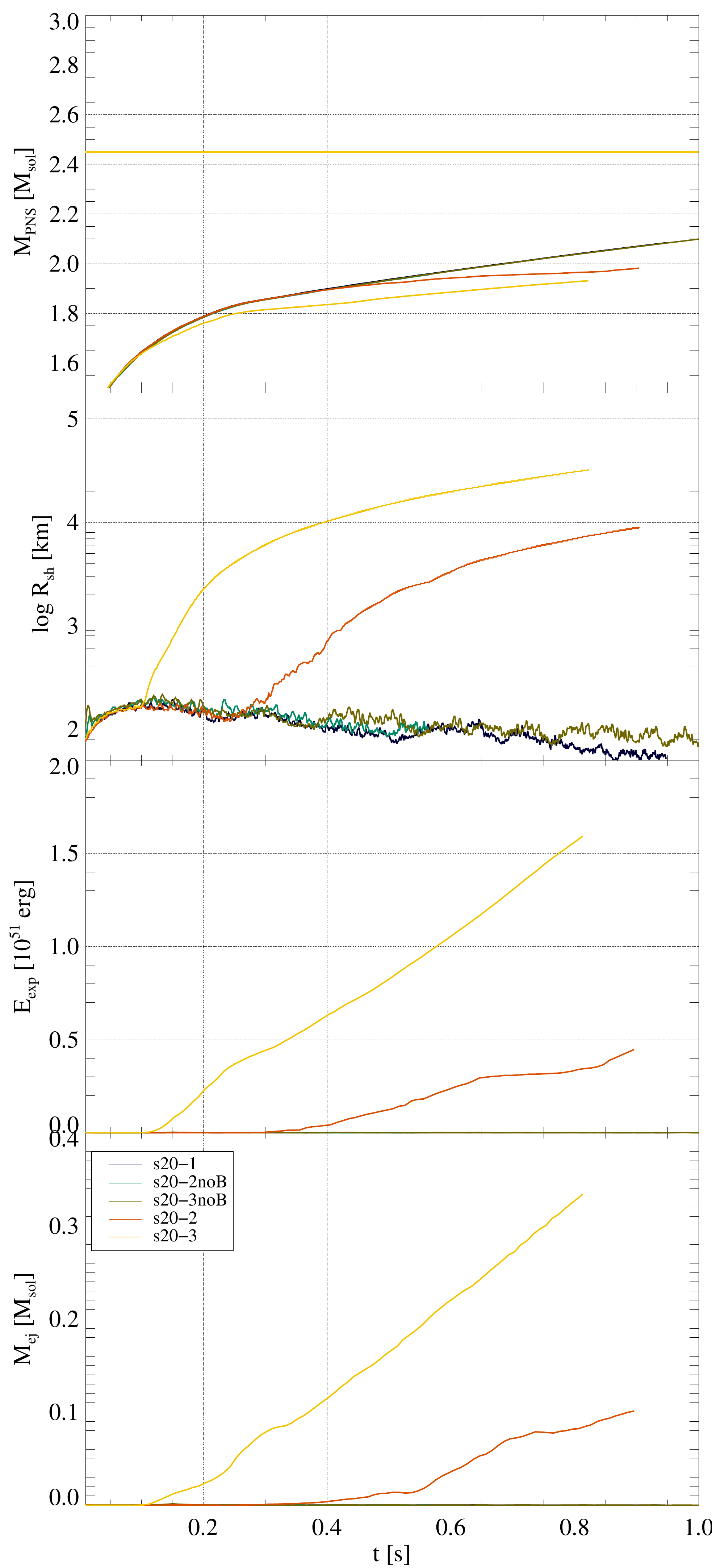}
  \caption{
    From top to bottom: 
    time evolution of the PNS mass, maximum shock radius, diagnostic
    explosion energy, and mass of the ejecta of models based on the
    progenitor \modelname{s20}. The maximum mass of
    non-rotating and cold neutron stars supported by
    the SFHo EoS employed for these models is annotated with a
    horizontal yellow line in the top panel. Since models s20-1,
    s20-2noB and s20-3noB do not explode, they do not appear in the
    lower two panels.
    }
  \label{Fig:s20-glob1}
\end{figure}

\begin{figure}
  \centering
  \includegraphics[width=0.95\linewidth]{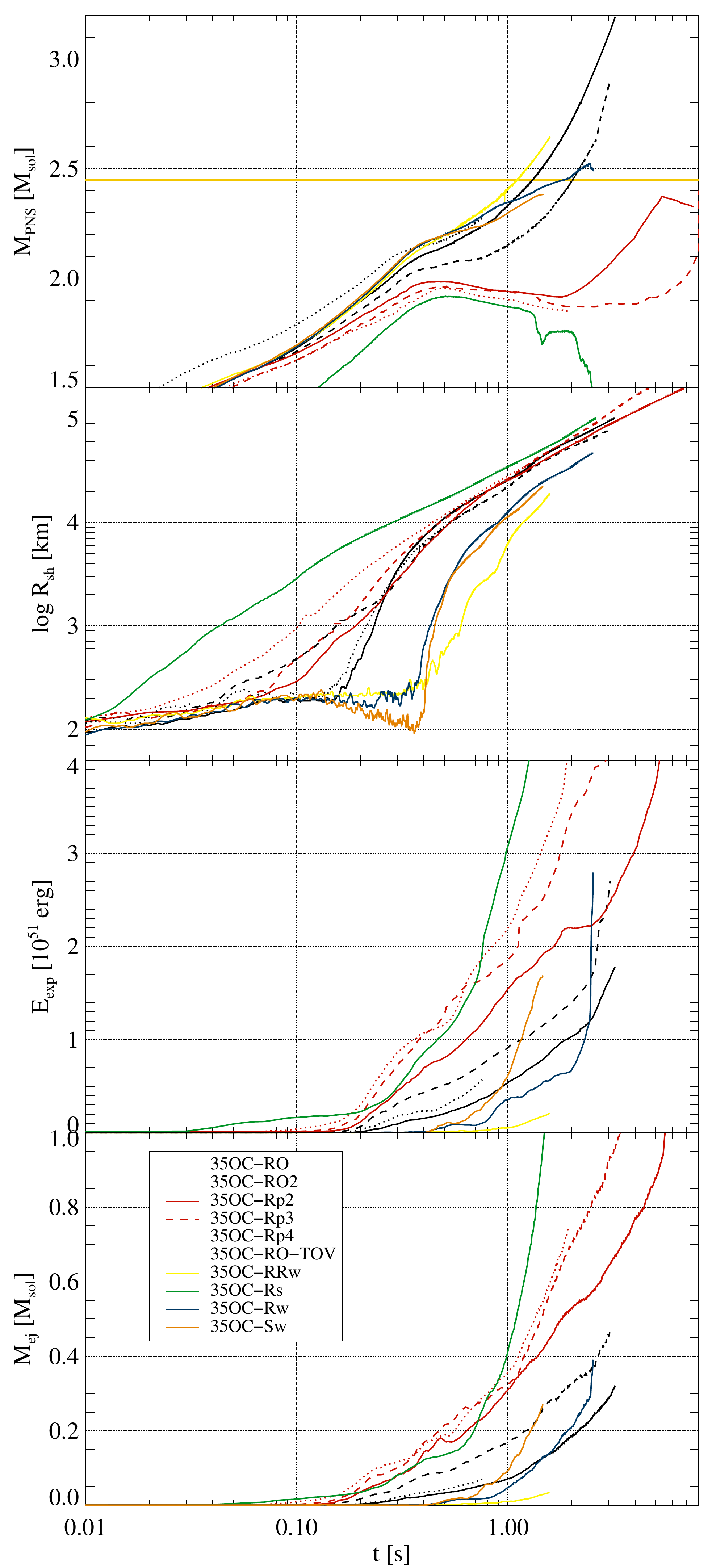}
  \caption{
    Same as \figref{Fig:s20-glob1}, but for models based on the
    progenitor \modelname{35OC}. 
  }
  \label{Fig:35OC-glob1}
\end{figure}

\begin{figure}
  \centering
  \includegraphics[width=0.95\linewidth]{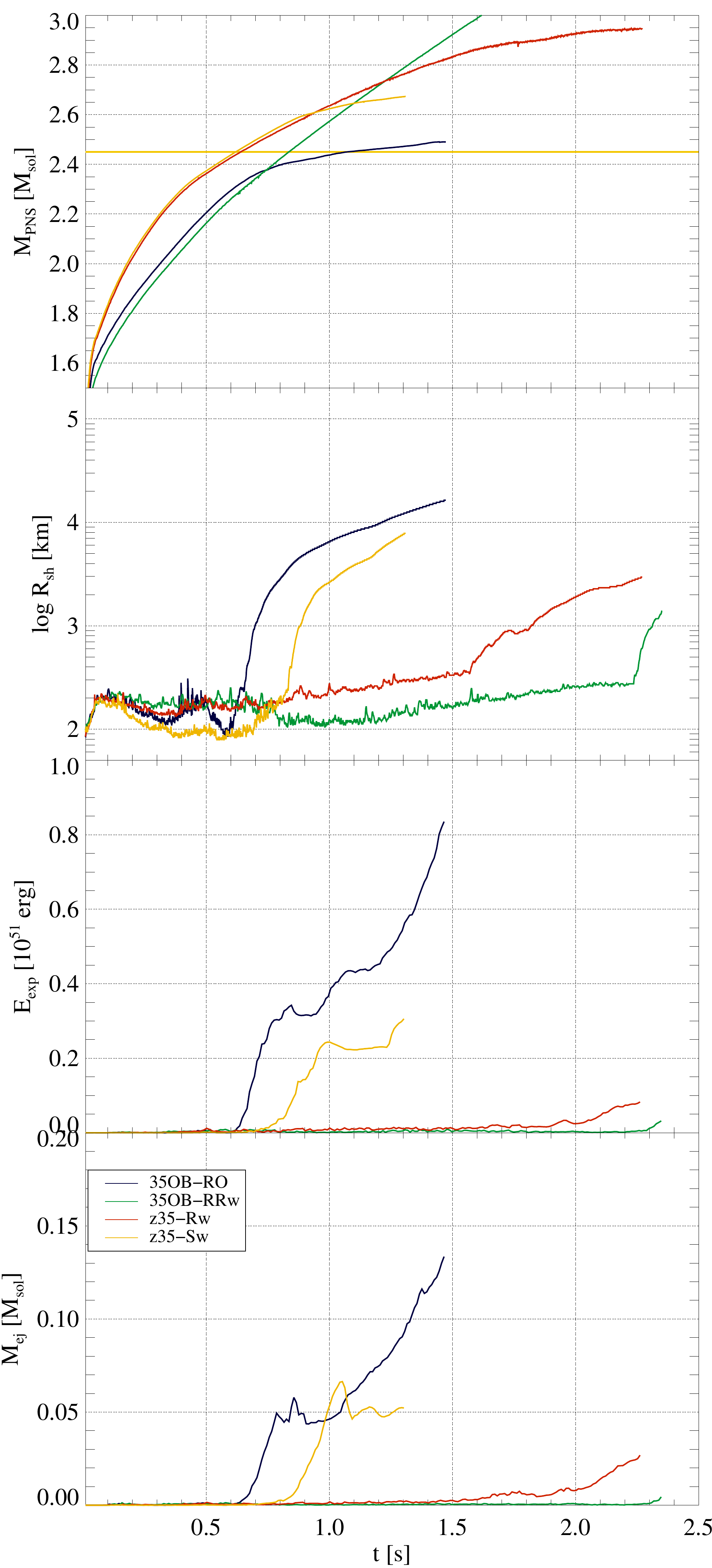}
  \caption{
    Same as \figref{Fig:s20-glob1}, but for models based on the
    progenitors \modelname{35OB} and \modelname{z35}.
  }
  \label{Fig:z35-glob1}
\end{figure}

\input{explprop-table}

Owing to their wide range of progenitor conditions, we find among our
models very distinct evolutionary paths.  Where they fall in this
characterisation is summarised in \tabref{Tab:models}.  The path of a
given model is not a function of the progenitor mass only, but is
crucially affected by the detailed structure, in particular the
compactness of the core
\citep[see][]{OConnor_Ott__2011__apj__BlackHoleFormationinFailingCore-CollapseSupernovae},
its rotational energy, and its magnetisation
\citep{Obergaulinger_Aloy__2017__mnras__Protomagnetarandblackholeformationinhigh-massstars}.
Varying only the rotational profile or only the magnetic field
strength can completely change the evolution of a core.  Of course, it
should be noted that while a variation of only one property of a
stellar model is possible in the idealised setting of a numerical
study, it would in reality entail a major adaption of all aspects of
the structure of the star.  Hence, some of our modifications are a bit
artificial and, thus, should not be treated as predictions for the
evolution of a particular star of a given mass and rotational and
magnetic energy but rather as a parameter study to tackle the impact
of individual processes. Besides, the progenitors we employ are the
result of one-dimensional stellar evolution models, which incorporate
rotation and the dynamical effects of magnetic fields with a limited
accuracy. In most cases, these models employ parameterizations of key
effects such as mass-loss or the braking action of magnetic
fields. Relatively small variations in these parameters result in
changes of the pre-collapse magnetic field and angular velocity
distribution, which we attempt to mimic with our parametrized
variations of these structural properties.

We show an overview of some of the most important global
quantities characterising the evolution of the models (PNS mass,
maximum shock radius, diagnostic explosion energy, ejecta mass)
in Figs.\,\ref{Fig:s20-glob1}--\ref{Fig:z35-glob1}.  Furthermore,
\tabref{Tab:Exprops} summarises properties of the models at the point
of explosion.  

\subsubsection{Neutrino-driven explosions}

Some of our models achieve shock revival by neutrino heating aided by
non-spherical gas flows.  In some of them, rotation causes important
modifications \wrt to the standard scenario for SN.  Examples for the
first (purely neutrino driven) and second (neutrino-rotationally
driven) scenario are \modls{35OC-Sw} and \modelname{35OC-Rw},
respectively.

\begin{figure}
  \centering
  \includegraphics[width=\linewidth]{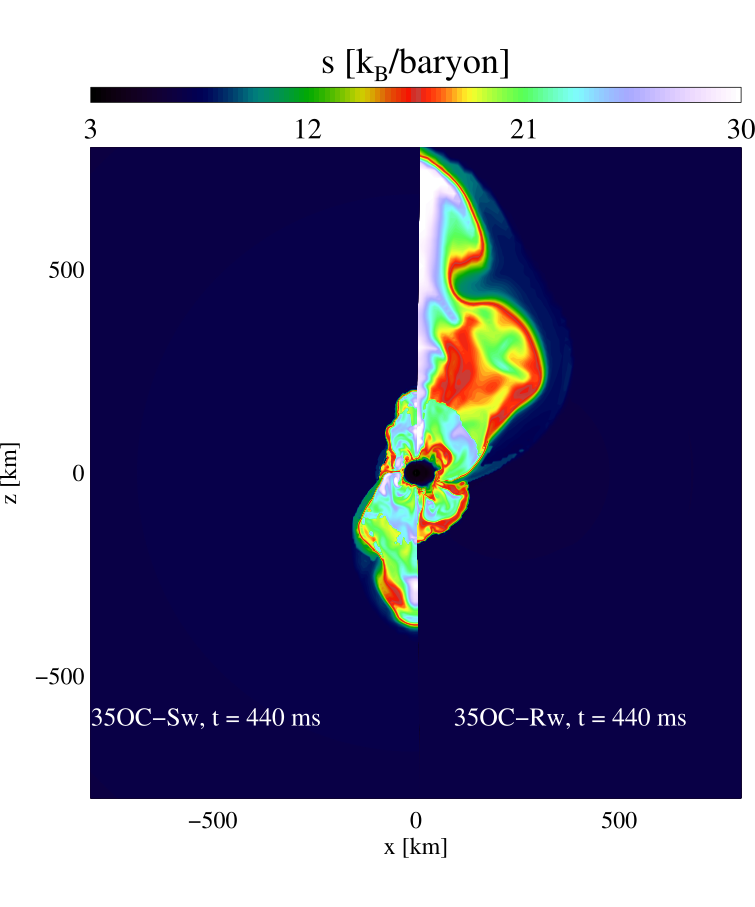}
  \caption{
    Comparison of the distribution of the specific entropy of
    \modls{35OC-Sw} and \modelname{35OC-Rw} at $t = 440 \, \ms$.
  }
  \label{Fig:35OC-RwSw-biplot}
\end{figure}

Most models launch an explosion within a few hundred milliseconds
post-bounce after a phase of pronounced activity of hydrodynamic
instabilities in the gain layer (\tabref{Tab:Exprops}).  Based on the
predominance of structures of angular extents of several tens of
degrees in the post-shock flow as well as in the deformations of the
shock surface, we characterise the models as dominated by convection
rather than by the standing accretion shock instability (SASI).
Magnetic fields are amplified in the gain layer and in the PNS, but do
not grow sufficiently as to affect the shock revival.  For both
moderately and rapidly rotating models, the supernova shock wave
starts to expand at high latitudes and the explosions take the form of
wide uni- or bipolar outflows along the symmetry axis
(\figref{Fig:35OC-RwSw-biplot}), into which a fraction of the matter
falling onto the PNS is redirected, mostly through downflows at low
latitudes.  The precise geometry of the downflows as well as the
angular width of the outflows and their fluxes of energy and mass can
fluctuate strongly with time.

Models of this class display a wide range of (diagnostic) explosion
energies and ejecta masses.  Most have energies a bit less than the
canonical SN explosion of $\zehn{51} \, \erg$, but a few
(\modls{35OC-Sw} and \modelname{35OC-Rw}; see
Fig.\,\ref{Fig:35OC-glob1}) exceed that value by the end
of the simulation and the growth of the energy indicates that others
(\eg \modls{s20-2}, Fig.\,\ref{Fig:s20-glob1}, and
\modelname{35OB-RO}, Fig.\,\ref{Fig:z35-glob1}) are likely to
do so shortly after the end of the simulations.

We note that a successful explosion and the formation of a BH are not
mutually exclusive.  Downflows increasing the PNS mass may coexist
with gas ejected in a very asymmetric manner.  Hence, we encounter BH
formation among exploding models such as \modl{35OC-Sw}.  A discussion
of the possible connection of such a scenario to the engines of long
GRBs within the collapsar model as well as the activity of
proto-magnetars as an alternative route to GRBs is performed in a
companion study (Paper II).

Finally, although successful explosions are the most common outcome,
several models, \eg \modl{s20-1/2noB/3noB}, 
fail to achieve shock revival within the simulation time. The failure,
if definite, will ultimately lead to the collapse of the PNS to a
BH, similar to the findings of
  \cite{Pan_et_al__2018__apj__EquationofStateDependentDynamicsandMulti-messengerSignalsfromStellar-massBlackHoleFormation,Chan_et_al__2018__apjl__BlackHoleFormationandFallbackduringtheSupernovaExplosionofa40MsunStar}\footnote{We
note that the final collapse of the PNS to a BH may be affected by the
choice of EOS \citep[see][]{Aloy_et_al__2019__mnras__Neutronstarcollapseandgravitationalwaveswithanon-convexequationofstate}}.
In a few cases, we could follow the evolution long enough to reach
this point, while in others the PNS mass grows too slowly for collapse
to occur within the simulation time of up to more than 2 seconds.  As
the sequence of models with the progenitor \modelname{s20} shows,
shock revival may fail at slow as well as rapid rotation.  In the
latter case, the reduction of the accretion luminosity due to very
deformed PNSs can be a decisive factor in avoiding an explosion.

The extremely rapidly rotating \modls{35OB-RRw} and \modelname{z35-Rw}
(see \figref{Fig:z35-glob1}) constitute cases at the boundary between
successful shock revival and failed explosions.  Their very high
rotational energy causes them to develop extremely oblate cores whose
\nusps extend to more than 100\,km in the equatorial plane.  As a
consequence, the gas falling towards the centre settles down at
comparably high radii, releasing less gravitational binding energy.
The neutrino luminosity is, hence, smaller than in models with less
flattened PNSs, which reduces the prospects of neutrino-driven shock
revival.  Their shock waves start to expand rather gradually at late
times (more than a second after bounce) and the post-shock gas
achieves comparably small positive energies around
$\mathcal{O}(10^{50} \, \erg)$.  Whether these energies are sufficient
for the shock to reach the stellar surface after the potential
collapse of the very massive PNSs to BHs is unclear, and would require
much longer computational times to be assessed.

\subsubsection{Magnetorotationally driven explosions}

Various processes can amplify the seed magnetic field of the core such
as compression by the radial flow, differential rotation, and
hydromagnetic instabilities with the MRI potentially playing a very
important role.  We defer a detailed discussion of its development to
a follow-up study focusing on the processes in the PNS and, for the
current analysis, start with the observation that several models
develop strong magnetic fields by a combination of the aforementioned
effects.

A sufficiently strong magnetisation may lead to explosions driven by
Maxwell stresses (with rotational contributions of different degree of
importance), which can set in earlier than their essentially
non-magnetic counterparts and may considerably exceed those in terms
of the explosion energies, such as in \modls{35OC-Rs},
\modelname{35OC-RO2}, \modelname{35OC-Rp2},
\modelname{35OC-Rp3}, \modelname{35OC-Rp4}, and \modelname{35OC-RO}
compared to the weaker magnetised version, \modelname{35OC-Rw}. Also
\modelname{s20-3} belongs to this set of models.

Even more than for weakly magnetised models, bipolar explosion
morphologies are characteristic for strongly magnetised ones.  In
particular the most intense magnetic fields are able to accelerate
collimated jets with moderately relativistic flow speeds of up to
$v^{} \lesssim c/3$.  The explosions tend to be the more energetic the
stronger the magnetic field is, with \modl{35OC-Rs} as the most
violent explosion reaching a diagnostic explosion energy of
$E_{\mathrm{exp}} \gtrsim \zehnh{4}{51} \, \erg$ within less than one
second.  The strong explosions partially suppress the accretion of gas
onto the PNSs, which therefore grow slower than for weaker magnetic
fields.  The suppression is most evident for \modls{35OC-Rs} and
\modelname{35OC-Rp4}, where the PNS mass ceases to grow after
$t \sim 500 \, \ms$ (Fig.\,\ref{Fig:35OC-glob1}).  \Modls{35OC-Rp2}
and \modelname{35OC-Rp3} develop a local maximum at about the same
time ($t \sim 500 \, \ms$), but then they grow again after
$t\sim 2\,$s. In the case of \modl{35OC-Rp2} a second local maximum
($M_\pnss \sim 2.35\msol < \Mmax$) is reached at $t \sim 5\,$s
postbounce, while \modl{35OC-Rp3} displays an ongoing PNS mass growth
by the end of the computed time ($t\sim 7\,$s). The behaviour of
\modl{35OC-Rs} after the maximum is explained in terms of the
morphological changes that the PNS undergoes. It changes its shape
from a prolate ellipsoid to a toroid at $t\sim 1.5\,$s (note the local
minimum in the green line of the upper panel of
Fig.\,\ref{Fig:35OC-glob1}). These two configurations display
morphologies akin to the A and C types found by, e.g.
\cite{Studzinska_et_al__2016__mnras__Effectoftheequationofstateonthemaximummassofdifferentiallyrotatingneutronstars}
for differentially rotating polytropes, respectively. Indeed our
results suggest that transitions between different types of
differentially rotating {\it quasi-}equilibrium models may be produced
as a result of the accretion/ejection of mass onto/from the PNS
outermost layers. Note that the strong decrease of the $M_\pnss$ in
\modl{35OC-Rs} is possible as a combination of two facts. Firstly, the
angular momentum redistribution resulting from the magnetic stress
acting on this extremely magnetized initial configuration and, second,
because these magnetic fields revert the accretion of mass onto the
PNS, yielding a mass ejection (see the fast increase of the ejecta
mass after the local maximum of $M_\pnss$ is reached;
Fig.\,\ref{Fig:35OC-glob1}).

\subsection{Elements of the dynamics}
\label{sSek:ResDe}

In order to understand the dynamics, we will explore how
various processes shaping their evolution differ across the range of
initial models.

\subsubsection{Mass accretion}

\begin{figure}
  \centering
  \includegraphics[width=\linewidth]{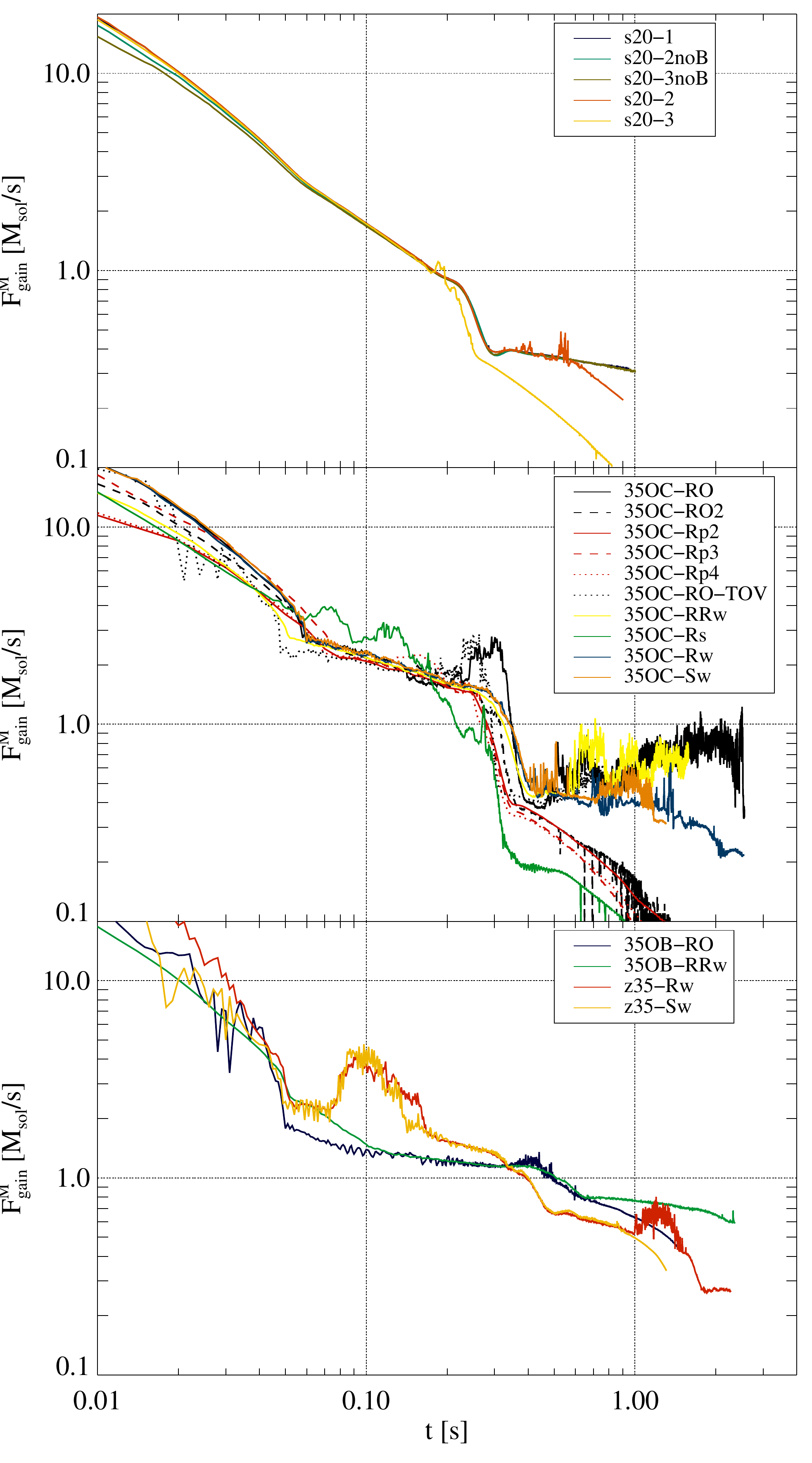}
  \caption{
    Mass flux through the shock wave as a function of time.  The
      three panels show, from top to bottom, the models based on progenitors
      \modelname{s20}, \modelname{35OC}, and
      \modelname{z35} and \modelname{35OB}. 
  }
  \label{Fig:mflxgain}
\end{figure}

We present the evolution of the total, \ie angularly integrated, mass
flux through the shock wave, $\dotMsh$, of all models in
\figref{Fig:mflxgain}.  The variations in this quantity reflect the
structure of the progenitor star. The more or less sudden drops in
$\dotMsh$ correlate with the different shells of the
progenitor star and the interfaces between them.  A further decrease
can be noted after the onset of an explosion as the shock wave
propagates out into shells of lower density and encounters a reduced
mass flux.

Our models exhibit the following values of $\dotMsh$
after the accretion of the first stellar interface: for \modl{s20-1},
we find $\dotMsh \lesssim 0.3 \, \msol\,\sek^{-1}$ after
$t \approx 300 \, \ms$, for \modl{35OC-Sw},
$\dotMsh \lesssim 0.4 \, \msol\,\sek^{-1}$ between
$t \approx 1 \, \sek$ and the collapse to a BH, for \modl{35OB-RRw},
$\dotMsh \approx 0.7 \, \msol\,\sek^{-1}$ after
$t \approx 1.5 \, \sek$, and for \modl{z35-Sw},
$\dotMsh \lesssim 0.4 \, \msol\,\sek^{-1}$ after
$t \approx 1.1 \, \sek$.
\cite{Suwa_et_al__2016__apj__TheCriterionofSupernovaExplosionRevisited:TheMassAccretionHistory}
investigated the mass accretion history of an extended set of
progenitors, compared to which our models show fairly high values.
Their analysis, limited to models without magnetic fields, would
suggest that rather high neutrino luminosities are required to trigger
explosions in our models.  A first conclusion from this comparison is
that strong magnetic fields circumvent the condition on the neutrino
luminosities as they are able to launch an explosion at mass accretion
rates far exceeding those in which models with weak or vanishing
fields undergo shock revival.  The most prominent examples are
\modl{35OC-Rs}, and \modelname{35OC-RO2} exploding before the
accretion of the surface of the $Fe$-core at mass accretion rates
around $2 \, \msol\,\sek^{-1}$, but also \modls{s20-3}, and
\modelname{35OC-RO} start their explosions slightly before the strong
decrease in the mass accretion rate.  Compared to \modl{s20-3}, the
slower rotation and weaker magnetic fields of \modl{s20-2} seem to go
into the same direction, enabling an explosion at a phase in which the
other models of the same progenitor (\modls{s20-1},
\modelname{s20-2noB}, \modelname{s20-3noB}) fail to explode.

\subsubsection{Neutrino emission}

\begin{figure*}
  \centering
  \includegraphics[width=0.48\linewidth]{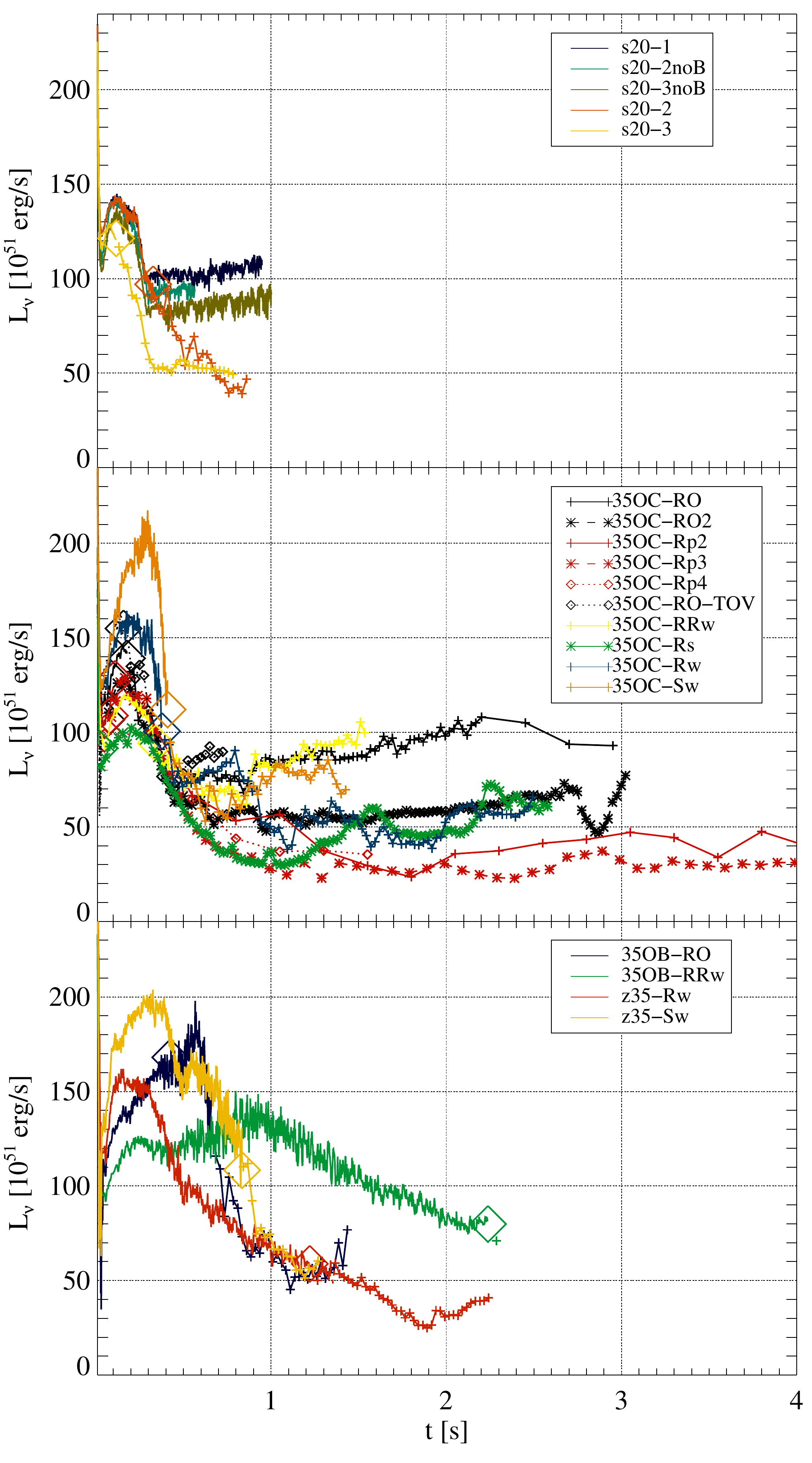}
  \includegraphics[width=0.48\linewidth]{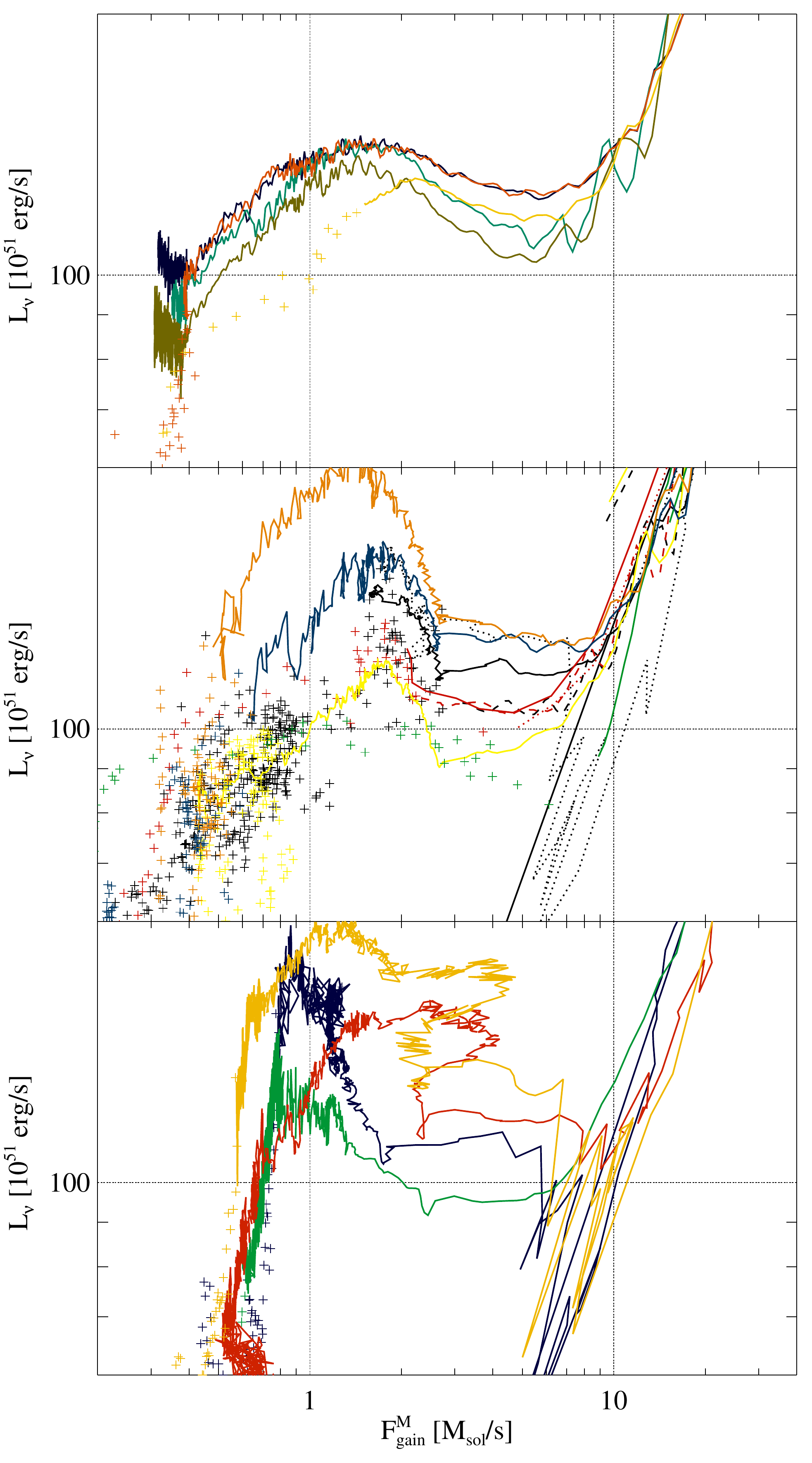}
  \caption{
    Sum of the electron neutrino and anti-neutrino luminosities as a
    function of time (left) and as a function of mass accretion rate
    through the shock wave (right).  In the right panels, solid lines
    and points represent the states of the models before and after the
    launch of an explosion, respectively. In the left panels,
    large rhombi indicate the onset of the explosion in each model.
  }
  \label{Fig:nlums}
\end{figure*}

\begin{figure*}
  \centering
  \includegraphics[width=0.48\linewidth]{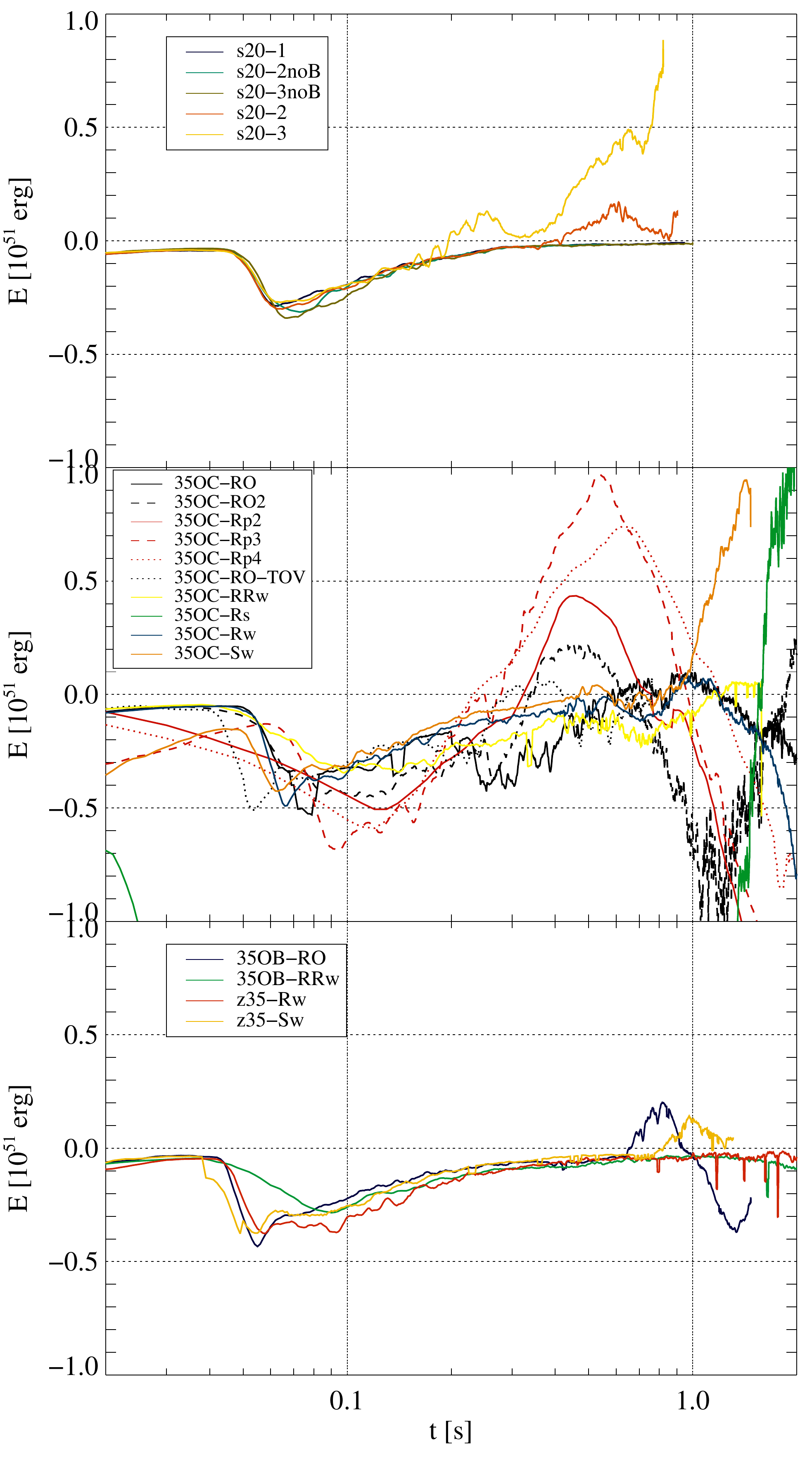}
  \includegraphics[width=0.48\linewidth]{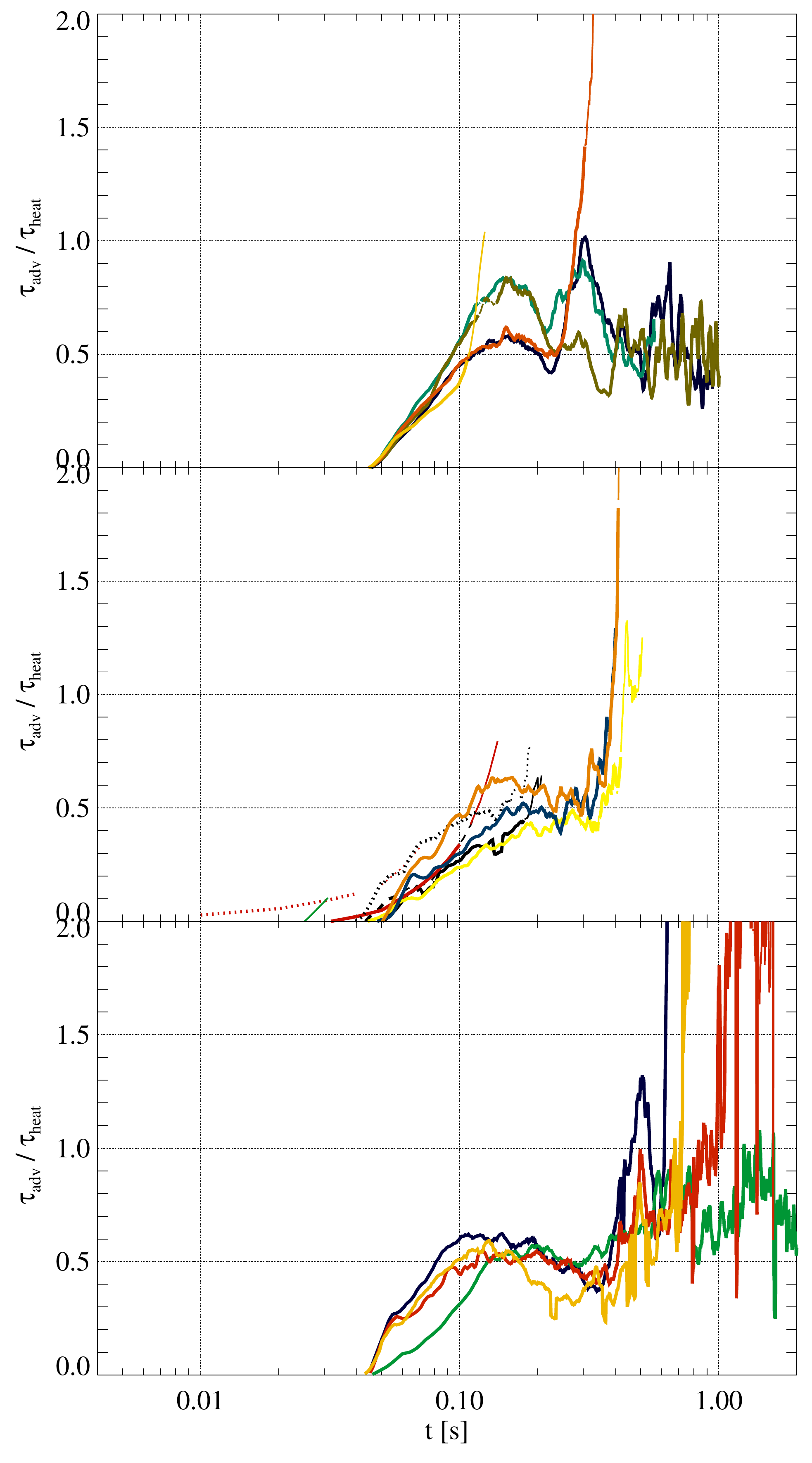}
  \caption{
    Time evolution of the total energy in the gain layer (left) and
    the ratio between advection and heating time scales in the gain
    layer (right panel).
  }
  \label{Fig:egaintau}
\end{figure*}

The time evolution of the total, \ie angle-integrated, neutrino
luminosities of our models, shown in the left panels of
\figref{Fig:nlums}, correlates with the dynamics in
several ways:
\begin{enumerate}
\item After the neutrino burst emitted at bounce, all neutrino
  flavours maintain high luminosities whose precise values depend on
  the mass accretion rate.  Models of progenitors \modelname{s20} (top
  left panel) and \modelname{35OC} (middle left panel), \eg exhibit
  high luminosities until $t \sim 250 \, \ms$ and $t \sim 400 \, \ms$,
  respectively.  After that, the decreases of $\dotMsh$ occurring when
  an interface between two shells falls onto the PNS correspond to
  notable decreases in the emission of neutrinos.
\item The variation of the luminosities among models of a series with
  the same progenitor, but different rotation and magnetic field can
  be relatively minor as in the group of models corresponding to the
  \modelname{s20} stellar progenitor or considerable as for model
  series \modelname{35OC}.
  In the former case, the main factor distinguishing
  between the neutrino light curves is the beginning of an explosion,
  which leads, by virtue of the decreased mass accretion rate, to a
  lower neutrino emission.  Among the models of group
  \modelname{35OC}, on the other hand, the one with slowest rotation,
  \modelname{35OC-Sw}, surpasses the neutrino luminosities of the
  rapid rotators with a delayed explosion (\modelname{35OC-Rw} and
  \modelname{35OC-RO}) by about $25 \, \%$ at $t \approx 200 \, \ms$.
  They in turn exceed the early exploding \modls{35OC-RO2} and
  \modelname{35OC-Rs} by about the same factor.  The inverse
  correlation between rotational velocity and neutrino emission is
  also observed for the other two groups of models.
\end{enumerate}

We combine the two quantities discussed so far and display the
evolution of the models in the phase space of mass accretion rate and
combined electron neutrino and anti-neutrino luminosities in the right
panels of \figref{Fig:nlums} similarly to previous studies
\citep{Burrows_Goshy_1993__apjl__ATheoryofSupernovaExplosions,Murphy_Burrows__2008__apj__CriteriaforCore-CollapseSupernovaExplosionsbytheNeutrinoMechanism,Nordhaus_et_al__2010__apj__Dimension_as_a_Key_to_the_Neutrino_Mechanism_of_CCSNe,Fernandez__2012__apj__HydrodynamicsofCore-collapseSupernovaeattheTransitiontoExplosion.I.SphericalSymmetry,Hanke__2012__apj__IsStrongSASIActivitytheKeytoSuccessfulNeutrino-drivenSupernovaExplosions,Janka__2012__ARNPS__ExplosionMechanismsofCore-CollapseSupernovae,Suwa_et_al__2016__apj__TheCriterionofSupernovaExplosionRevisited:TheMassAccretionHistory,Ertl_et_al__2016__apj__ATwo-parameterCriterionforClassifyingtheExplodabilityofMassiveStarsbytheNeutrino-drivenMechanism}.
As in general the mass accretion rate decreases with time, the models
traverse the diagrams from the right to the left.  Their trajectories
are represented by solid lines and symbols before and after the onset
of an explosion, respectively.  Hence, the point where a line ends
marks the conditions at the revival of the shock wave (henceforth,
revival luminosities and mass accretion rates).  A relation between
the mass accretion rate and a critical luminosity required for
launching an explosion should show up in the distribution of the end
points of the trajectories.

However, we do not find any evidence for any such direct relation.  On
the contrary, explosions seem to occur almost randomly across the
entire space of parameters.  Several models explode at moderate
$\dotMsh < 1 \, \msol \, \sek^{-1}$.  Within this group,
however, the revival luminosities differ significantly.  We point, \eg to
\modls{s20-2}, \modelname{35OC-Sw}, \modelname{z35-Sw}, and
\modelname{35OB-RO} with revival luminosities differing by a factor of
about two \tabref{Tab:Exprops}.  The connection between the
$\dotMsh$-$L_{\nu}$-trajectories and the actual
evolution of the models is complicated further by the fact that some
models fail to explode at values of $\dotMsh$ and
$L_{\nu}$ that correspond to the shock revival in others.  We refer
to, \eg \modls{s20-1} and \modelname{s20-2noB} that, in contrast to
\modl{s20-2} with a very similar trajectory, do not explode when
reaching values around $\dotMsh \approx 0.4 \, \msol \,
\sek^{-1}$ and $L_{\nu} \approx \zehn{52} \, \erg \,\sek^{-1}$.

The consideration of strong magnetic fields further complicates
finding a distinction between exploding and non-exploding models in
this space of parameters.  The early explosions of, \eg
\modls{35OC-Rs/O/O2} or \modelname{s20-3} occurs at large
$\dotMsh$.  While the revival luminosities are large as
well, they are exceeded by the values of non-exploding models at the
same mass accretion rates.

To view the interplay of the two processes of accretion and heating by
neutrinos from a different perspective, we consider the time scales
for advection through the gain layer, $\tauadv$, and heating,
$\tauhtg$ \citep[see, \eg
][]{Janka__2001__aap__Conditionsforshockrevivalbyneutrinoheatingincore-collapsesupernovae,Thompson_Quataert_Burrows__2004__ApJ__Vis_Rot_SN,Murphy_Burrows__2008__apj__CriteriaforCore-CollapseSupernovaExplosionsbytheNeutrinoMechanism}.
The former is given by $\tauadv = D / |\langle v^r \rangle|$, where
$D$ and $\langle v^{r} \rangle$ are the radial extent of the gain
layer and the volume average of the radial velocity of the gas inside
the gain layer, respectively.  An alternative definition for the
advection time scale can be the ratio of the mass in the gain layer
and the mass accretion rate through the shock wave
\citep[e.g.][]{Summa_et_al__2016__apj__Progenitor-dependentExplosionDynamicsinSelf-consistentAxisymmetricSimulationsofNeutrino-drivenCore-collapseSupernovae,OConnor_Couch__2018__apj__Two-dimensionalCore-collapseSupernovaExplosionsAidedbyGeneralRelativitywithMultidimensionalNeutrinoTransport}.
For averages over the entire gain layer, our formulation yields
results that are equivalent to the alternative definition.  However,
the generalisation of the version based on the flow speeds to a
multi-dimensional form seems more straightforward than a formula
involving the total mass in the gain layer, which is a global rather
than angle-dependent quantity.  This property motivated us to adopt it
rather than the one based on the masses and mass flows.

We define $\tauhtg$ as the time required for
neutrinos depositing energy at a rate $Q_{\nu}$, averaged over the
entire volume of the gain layer, $V_{\rm gain}$, \ie
\begin{equation}
  \label{Qnu}
  Q_{\nu} (t) = \frac{1}{V_{\rm gain}}\int_{V_{\rm gain}} \alpha Q_\star^0(t) dV,
\end{equation}
to raise the total energy
(gravitational plus MHD energy including internal, magnetic, and
kinetic contributions) of the gas in the gain layer,
$\Egain = M_{\mathrm{gain}} \langle \phi \rangle +
E^{\mathrm{mhd}}_{\mathrm{gain}}$, above zero ($\langle\phi \rangle$
is the average gravitational potential and $M_{\mathrm{gain}}$ is the
mass contained in the gain layer), \ie
\begin{equation}
  \label{Gl:tauheat}
  \Egain (t_0) = - \int_{t_0}^{t_0+\tauhtg} Q_\nu (t) dt.
\end{equation}
One-dimensional models undergo shock revival if neutrino heating is
faster than advection, \ie if $\tautau > 1$.  This criterion is also
applicable in multi-dimensional models, where deviations from
spherical symmetry such as hydrodynamic instabilities increase the
advection time, making an explosion more likely than in spherical
symmetry.

We show the evolution of the total energy and the ratio between the
two time scales in \figref{Fig:egaintau} (see also
\tabref{Tab:Exprops}).  For the non-exploding models of the
\modelname{s20} group, \ie \modelname{s20-1}, \modelname{s20-2noB},
and \modelname{s20-3noB}, $\Egain$ never becomes positive and heating
is usually slower than advection.  Though $\tautau$ may occasionally
reach unity, it tends to oscillate around moderate values $\sim 0.5$.
For \modl{s20-3}, the
explosions set in about $100\, \ms$ before $\Egain$ gets positive when
the timescale ratio is around $\tautau \approx 0.45$, \ie at a value
which in other models does not suffice for shock revival.  Shock revival
also coincides with a rapid increase of $\tautau$.  It is, however,
problematic to unambiguously define the onset of the explosion.  The
shock starts to expand at $t \approx 250 \, \ms$ at the south pole.
The basic structure from which the explosion emerges, \viz a strongly
magnetised polar region, is then already formed.  Its presence might
be an argument for identifying the onset of the explosion with this
point, even though the shock propagates outwards rather slowly at
first.  
The subsequent increase of the timescale ratio, which exceeds 1 when
the maximum shock radius is $R_{\mathrm{sh; max}} \approx 150 \, \km$,
is caused mostly by a rising advection timescale, \ie by the very
shock expansion itself.  The heating timescale, on the other hand,
follows during this phase closely that of the non-exploding
\modls{s20-1} and \modelname{s20-2noB}.  We take this observation as
an indication that the growth of $\tautau$ is a consequence of the
beginning of the explosion rather than its origin.

Likewise, the shock revival of models with the progenitor
\modelname{35OC} may precede the moment in which the entire gain layer
achieves positive total energies by several 100\,ms.  Their explosions
cannot be characterised by a critical value of $\tautau$.  Shock
runaway may occur at values as low as $\tautau \approx 0.1$
(\modelname{35OC-RO2}).
If the explosions discussed so far all occur at $\tautau < 1$,
\modls{35OC-Sw} and \modelname{z35-Sw} do not launch an explosion
until $\tautau$ has grown considerably beyond unity, which happens
well before reaching $\Egain>0$.

In \modls{35OC-Sw}
and \modelname{z35-Sw}, the onset of the shock runaway
also coincides with an increase of the ratio of the heating and
advection timescales.  Triggered by the drop of the mass accretion
rate, the former model launches a failed attempt to explode that
culminates at $t \approx 500 \, \ms$, during which $\tautau$ exceeds
unity.  The gain layer energy, however, remains negative during this
stage.  Only afterwards, a sustained shock expansion starts.  When the
maximum (\ie polar) shock radius begins to grow, we find
$\tautau \approx 1$.  The shock expansion picks up speed very quickly
while $\tautau$ grows beyond unity.  \Modl{z35-Sw} develops a brief
shock expansion at $t \approx 500 \, \ms$, again caused by the lower
mass accretion rate, leading to $\tautau \simeq 1$, which,
however, is insufficient for an explosion.  In this model, the outflow
launching happens at $t \approx 800 \, \ms$, around the same time as
$\Egain$ changes sign and $\tautau>24$.

The final two models, extreme rotators both, behave very differently
from the ones discussed so far. Except for its very final stage,
\modl{35OB-RRw} maintains $\Egain < 0$ and $\tautau < 1$ both before the shock runaway and
even during the final rapid shock
expansion.  \Modl{z35-Rw} shows a similar evolution of the shock
radius, though with a faster expansion during the second phase and a
more gradual one at the end.  The gain layer remains always bound at
all times, but, in contrast to \modl{35OB-RRw}, $\tautau > 1$ after $t
= 1 \, \sek$.

We are, thus led to the conclusion that a examination of the global
mass accretion rates, luminosities, advection and heating times alone
does not yield a consistent picture of the conditions for an
explosion.  Finally, we point out that the magnetic effects and the
non-spherical geometry of the explosion are main factors in the
inconsistency of the explosions of these and similar models with the
global explosion conditions of mass accretion rate and neutrino
luminosity or advection and heating times.

\subsubsection{Rotation}

\begin{figure}
  \centering
  \includegraphics[width=0.98\columnwidth]{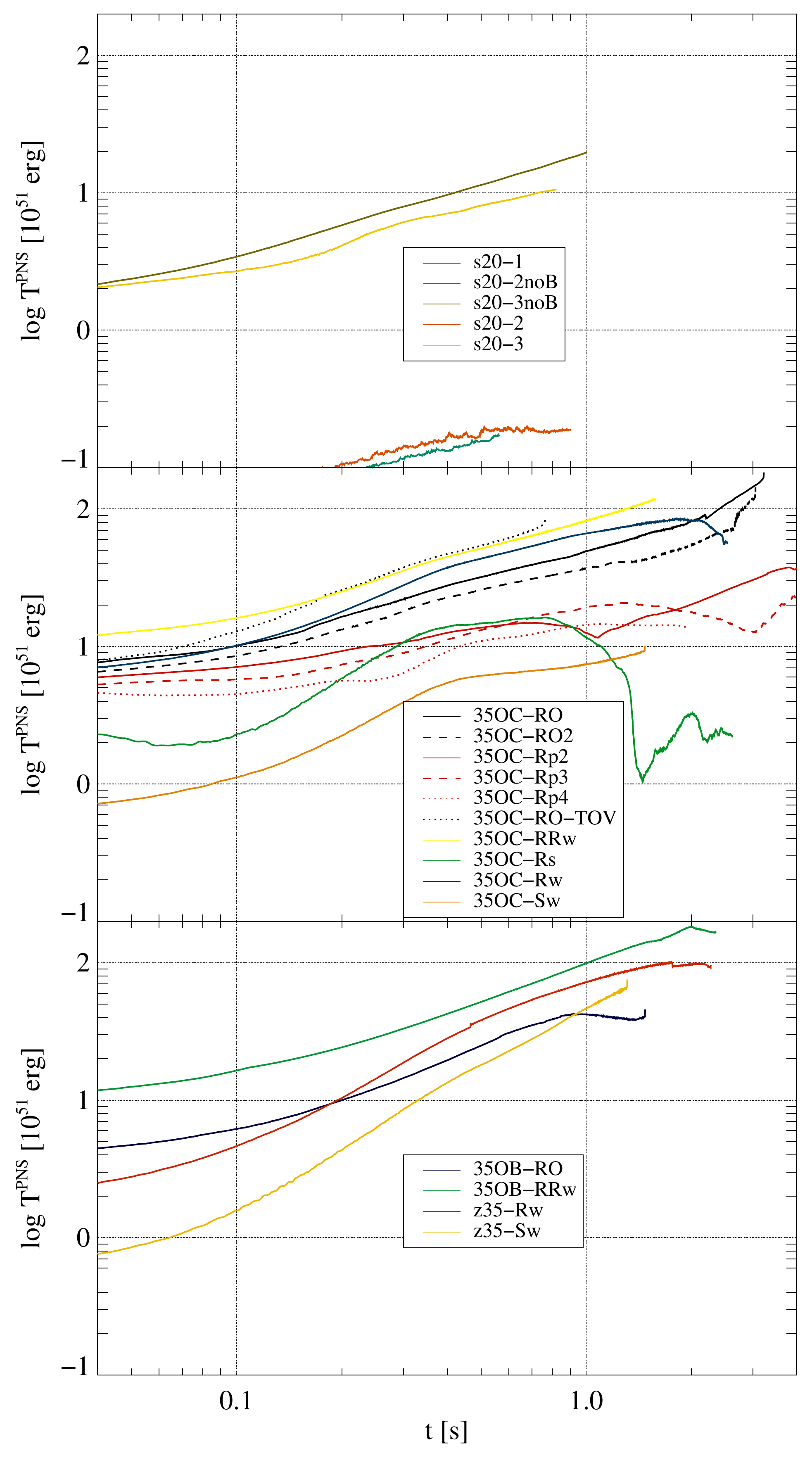}
  \caption{
    Time evolution of the rotational energy of the PNSs.
  }
  \label{Fig:erots}
\end{figure}

\begin{figure}
  \centering
  \includegraphics[width=0.98\columnwidth]{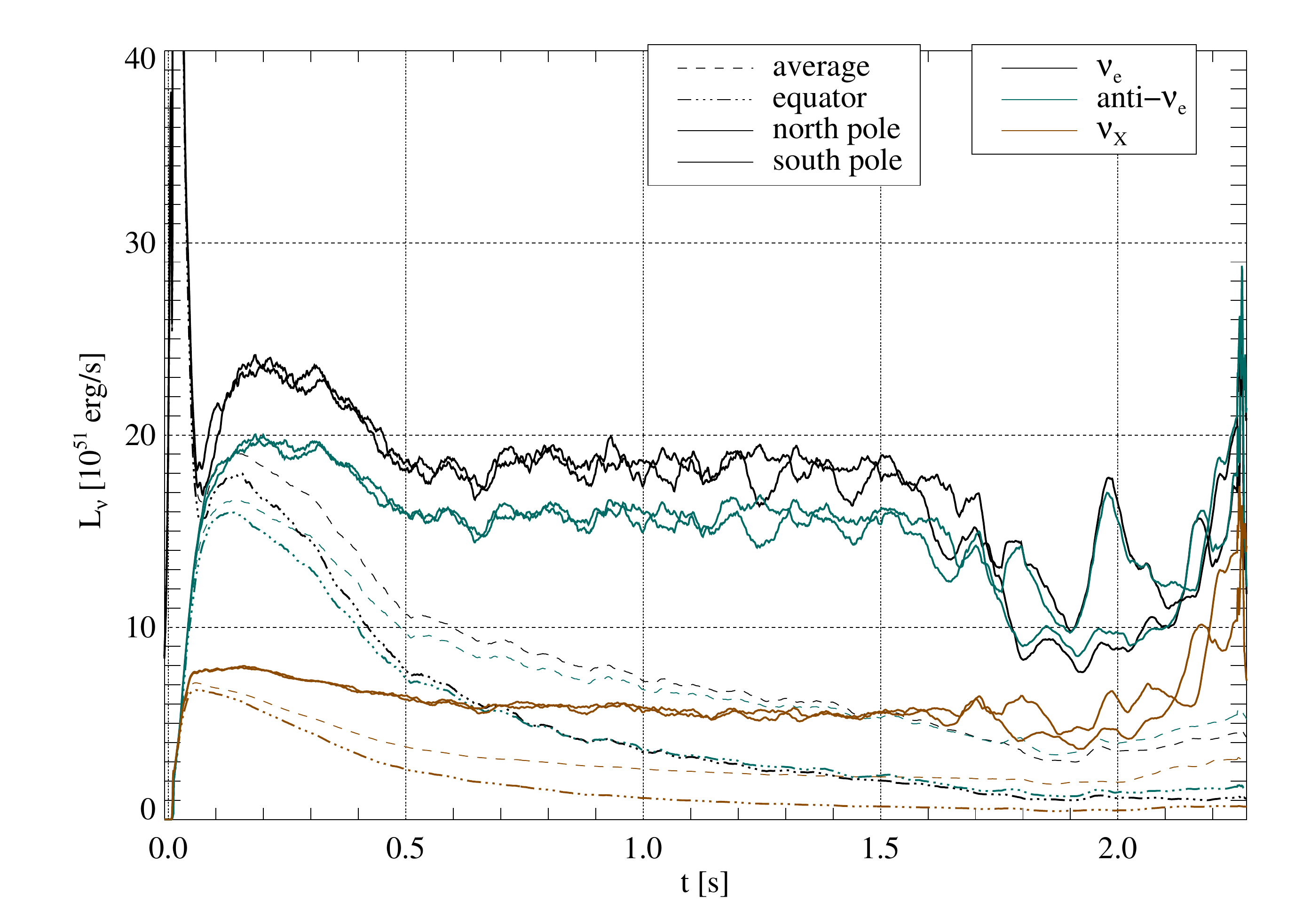}
  \caption{
    Time evolution of the isotropic equivalent neutrino luminosities
    of \modl{z35-Rw}.  Colours distinguish between flavours.
    Solid and dash-triple-dotted lines show the neutrino emission along the rotational
    axis and the equator, respectively, and  dashed lines represent the
    angularly averaged luminosities.
  }
  \label{Fig:z35-Rw-nlums}
\end{figure}

The radial infall during collapse and after bounce concentrates the
angular momentum of the progenitor in the central regions.
Consequently, the rotational energy of most PNSs of our models
increases beyond $\Erotpns > \zehn{52} \, \erg$
within a few 100\,ms after bounce (see \figref{Fig:erots}).  Such
high values represent a few per cent of the gravitational energy of
the PNS.  Though the models should avoid the parameter range of
dynamical bar-mode instability, we cannot rule out that they would
develop secular instabilities if the constraint of axisymmetry were
dropped.\footnote{It is, however, remarkable that none of the 3D
  (low-resolution) versions of the prototype models
  \modelname{35OC-RO} and \modelname{35OC-Rs} have developed these
  instabilities.} The exceptions are the initially slower rotators
\modls{s20-1}, \modelname{s20-2}, \modelname{s20-2noB}, and
\modelname{35OC-Sw}.

Mass accretion slows down at late times, in particular after the start
of an explosion.  Moreover, due to the negative radial gradient of the
angular velocity, each newly accreted shell of mass adds relatively
less rotational energy than its immediate predecessor.  The rate at
which the addition of rotational energy occurs, thus, depends on the
angular velocity profile of the shells, with outer layers, where the
gradients are steepest, contributing less.  Consequently, the angular
momentum of the PNS grows slower after a few hundred ms than early on.

Very high rotational energies, $\Erotpns \gtrsim \zehn{52} \, \erg$,
cause significant flattening of the PNS.  The most extreme case is
that of \modl{z35-Rw}.  At $t \approx 1.6 \, \sek$, the equatorial
layers of the PNS extend to around 100\,km at the equator compared to
less than 20 km at the poles.  Thus, matter accreted onto the core
ends up at high radii in an equatorial bulge.
  
The fast rotation induces a strong latitudinal variation of the
neutrino emission
\citep{Kotake_etal__2004__Apj__SN-magrot-neutrino-emission}.
In the most extreme cases (\modelname{35OB-RRw},
\modelname{z35-Rw}), the neutrino flux (or isotropic equivalent
neutrino luminosity) of electron-type flavours at the poles can exceed
that at the equator by a factor of more than 5.
We show the time evolution of the isotropic equivalent neutrino
luminosities of all flavours in \figref{Fig:z35-Rw-nlums}.  The
emission along the rotational axis remains very strong for a long
time, whereas the emission in the equatorial direction is much
weaker and gradually decreases.   
The neutrino flux
anysotropy helps breking the spherical symmetry and hence, contributes
to the success in the outflow launching
\citep{Obergaulinger_Aloy__2017__mnras__Protomagnetarandblackholeformationinhigh-massstars}.

The global rotational energies show a similar evolution in models with
the more complex rotational profiles given by stellar evolution
progenitors instead of the simpler $j-const$ profiles such as
\modls{35OC-Rw} and \modelname{35OB-RO}.  
Both models with the $\Omega$-profiles taken from the
stellar-evolution calculations, \modelname{35OB-RO}, and weak magnetic
fields, \modelname{35OC-Rw}, show moderate centrifugal flattening of
the PNS, at least during the first 1.5\,s after bounce.  The
rotational profile is, like in \modl{z35-Rw}, roughly cylindrical.

\subsubsection{Hydrodynamic instabilities}

\begin{figure}
  \centering
  \includegraphics[width=\linewidth]{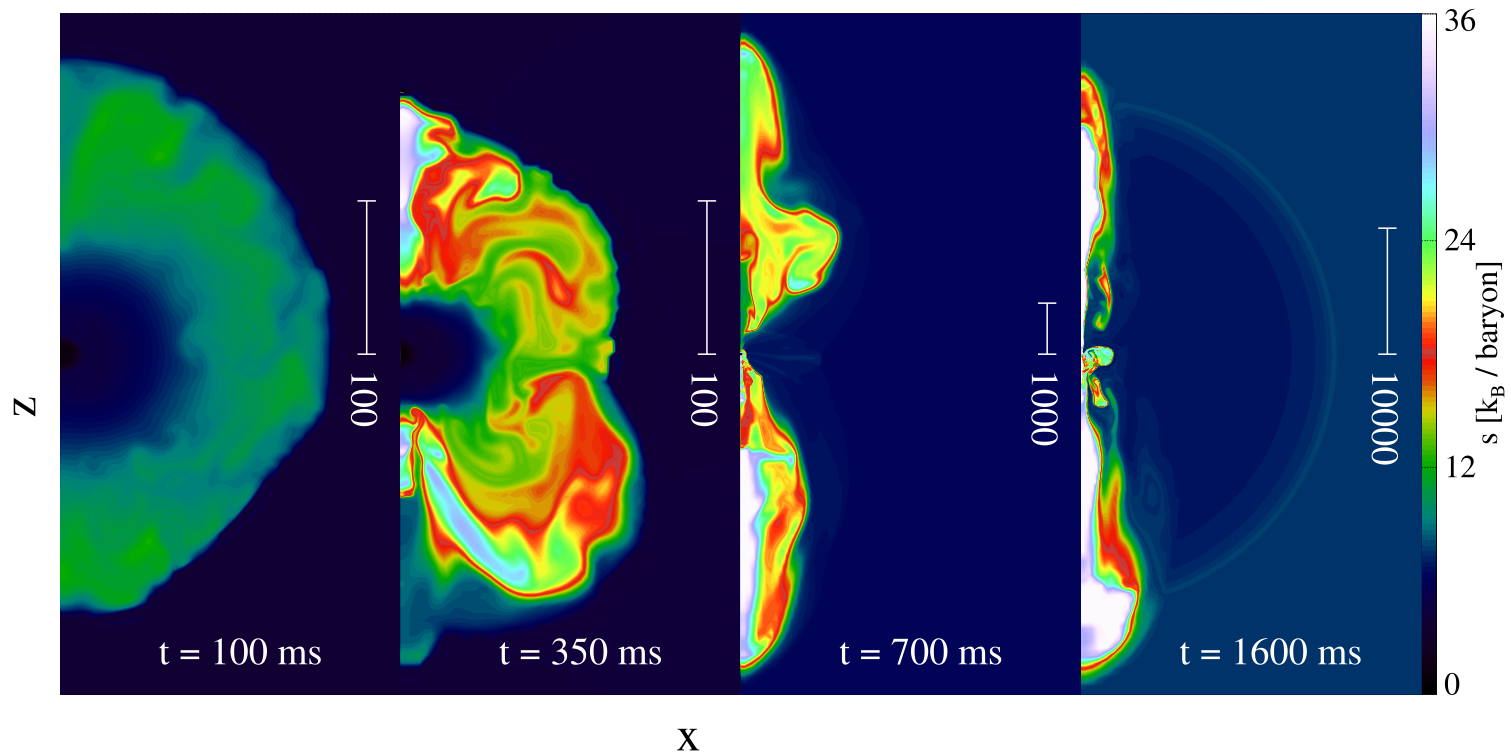}
  \caption{
    Distribution of the specific entropy of the core of \modl{35OC-Rw}
    at four times as indicated: $t = 100 \, \ms$, \ie, before shock revival, $t =350\,\ms$,
    \ie, around the onset of the explosion, and two times after that.
    The scale of each panel is shown by the ruler whose length in
    units of km is displayed.
  }
  \label{Fig:35OC-Rw-2d}
\end{figure}

\begin{figure}
  \centering
  \includegraphics[width=\linewidth]{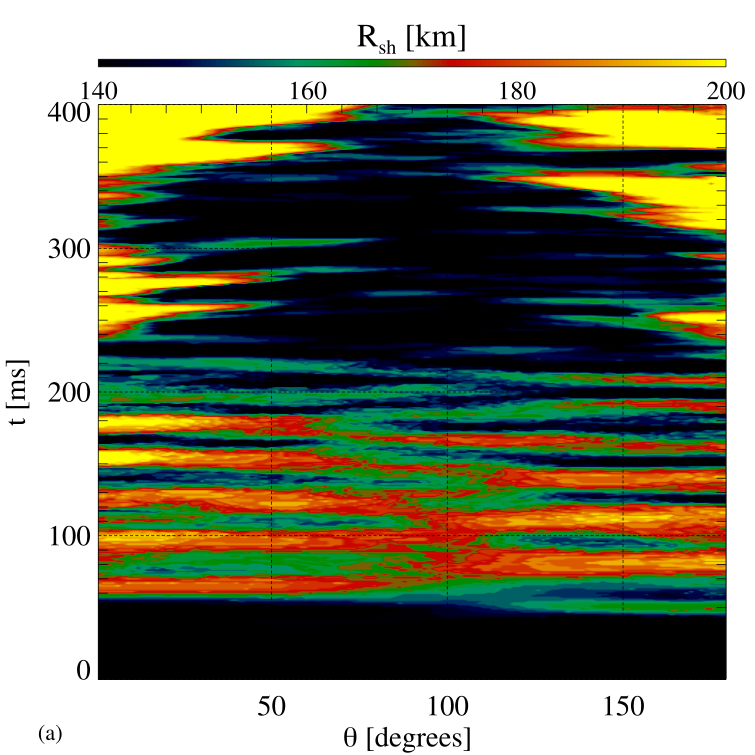}
  \caption{
    Evolution of the shock radius, $R_{\mathrm{sh}}$, of \modls{35OC-Rw} as a
    function of angular coordinate and time.
  }
  \label{Fig:35OC-Rw-shockrad}
\end{figure}

\begin{figure}
  \centering
  \includegraphics[width=\linewidth]{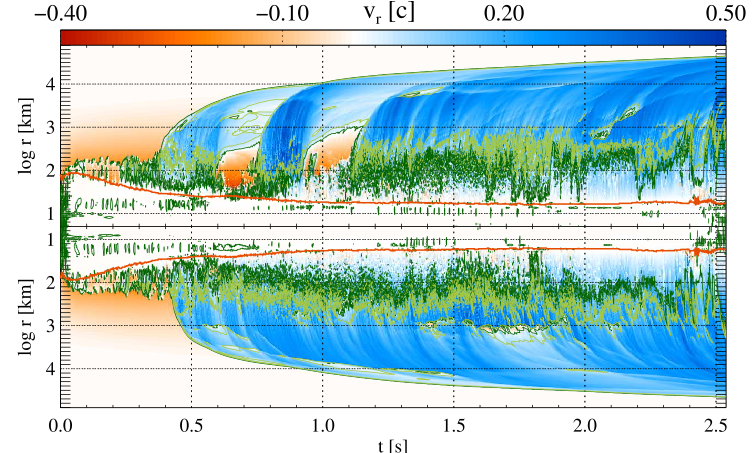}
  \caption{
    Radial velocity (in terms of the speed of light) of
    \modl{35OC-Rw} as a function of time and radius along the
    north and south pole corresponding to the upper and lower
    halfs of the panel, respectively. The red line marks
    the location of the electron \nusps. The green
    lines separate regions where the gas velocity is sub-Alfv{\'e}nic
    from ones where it is 
    super-Alfv{\'e}nic (dark green) and regions of sub-fast velocities
    from those of super-fast velocities (light green).  Note that
    close to the PNS the gas is typically sub-Alfv{\'e}nic and
    sub-fast and undergoes a transition to first super-Alfv{\'e}nic
    and then super-fast velocities at higher radii.
  }
  \label{Fig:35OC-Rw-jetvra}
\end{figure}

Though their impact should not be discarded, the magnetic fields are
not the main drivers in the explosions of \modls{s20-2},
\modelname{35OC-Sw}, and \modelname{35OC-Rw}.  The three models
develop explosions when the mass accretion rates have dropped
considerably, $\dotMsh < 1 \, \msol\,\sek^{-1}$ and at
different total neutrino luminosities.  The ratio between advection
and heating time scales at the beginning of shock runaway also differs
significantly between them (see \tabref{Tab:Exprops}). 
Most notably, even weak
magnetic fields significantly improve the collimation of the SN
ejecta, which adopts the typical geometry of bipolar jets at
sufficiently long times post bounce (see the two rightmost panels of
\figref{Fig:35OC-Rw-2d}).

Besides rotation, the gain layer is chiefly affected by hydrodynamic
instabilities creating meridional flows.  In \modl{35OC-Rw}, the
dominant modes have angular extents of several tens of degrees,
generating eddies of roughly equal radial and lateral extents (see
\figref{Fig:35OC-Rw-2d} at $t= 100 \, \ms$).  These eddies are thus
more typical for convective modes than for the SASI.  However, we also
find coherent north-south sloshing modes of the shock surface until
$t \approx 200 \, \ms$ in \modl{35OC-Rw} with large and small shock
radii oscillating between the north and south poles that show up in
the alternating pattern of blue-green and red-yellow colours in
\figref{Fig:35OC-Rw-shockrad}.

The pattern becomes weaker after $t \approx 200 \, \ms$, and the shock
recedes at the equator while it stabilizes at the poles until starting
to expand there rapidly at $t \approx 350 \, \ms$.  The SASI
oscillations coupling regions in an entire hemisphere (see, \eg the
large arc of hot gas extending from the polar cap of the PNS to
regions close to the equator in \figref{Fig:35OC-Rw-2d}) and, to a
lesser degree, the smaller-scale convective eddies lead to an enhanced
mixing of fluid elements across different latitudes.

While in non-rotating models a main effect of non-radial instabilities
lies in an increase of the dwell time in the neutrino-heating region,
we find here an opposite process.  The lateral motions induced by
convection and the SASI reduce the time a fluid element spends in the
polar region.  Consequently, they effectively reduce neutrino heating,
which achieves its highest efficiency at the poles and delays the
onset of the explosion until well after the two time scales are equal
locally.

As we show in \figref{Fig:35OC-Rw-shockrad}, the shock wave is highly
asymmetric both before and at the onset of the explosion.  Until
$t \sim 200 \, \ms$, the asymmetry is caused largely by $l=1$ sloshing
modes of the SASI, which then give way to a global pole-to-equator
asphericity as the shock is revived along the polar axis.  Shortly
before shock revival ($t \sim 350 \, \ms$), the shock wave has an axis
ratio around $5:3$.  Within the next 50 ms, the axis ratio increases
to $4:1$.  The asymmetry translates into pronounced latitudinal
variations of, \eg, pressure along the shock wave.
\cite{Nagakura_et_al__2013__apj__ASemi-dynamicalApproachtotheShockRevivalinCore-collapseSupernovae}
analysed the effect of such fluctuations on the conditions for shock
revival.  We can quantify the fluctuations in terms of the radial
velocity of the pattern speed of the shock front,
$v_{\mathrm{sh}} (\theta) = \partial_{t} R_{\mathrm{sh}} (\theta)$.
In the time leading up to the shock revival, we find variations of the
order of
$\delta v_{\mathrm{sh}} (\theta) = v_{\mathrm{sh}} (\theta) - \langle
v_{\mathrm{sh}} \rangle \sim 5 \times 10^8 \, \cms$, where
$\langle v_{\mathrm{sh}} \rangle$ denotes the angular average of
$v_{\mathrm{sh}} (\theta)$.  Expressed in terms of the post-shock
pressure $P_{\mathrm{sh}} \theta$ , the (analogously defined)
variation
$\delta P_{\mathrm{sh}} (\theta) = ( P_{\mathrm{sh}} (\theta) -
\langle P_{\mathrm{sh}} \rangle) / \langle P_{\mathrm{sh}} \rangle$
reaches values of $\delta P_{\mathrm{sh}} \sim 0.5$.  Such values are
consistent with the critical fluctuations for inducing shock expansion
found by
\cite{Nagakura_et_al__2013__apj__ASemi-dynamicalApproachtotheShockRevivalinCore-collapseSupernovae}.

As mentioned above, the presence of weak or moderate magnetic fields
in the stellar progenitor translates into the development of
collimated bipolar ejecta. This ejecta is modulated by the specific
dynamics of hydrodynamical instabilities which introduce a fairly
large degree of stochasticity over a long time and fluctuations of the
speed of the ejecta and the associated fluxes of energy and mass. As
\figref{Fig:35OC-Rw-jetvra} shows, the radial velocity immediately
outside the \nusp is more fluctuating than in \modl{35OC-RO} (which
contains an initially the same rotational energy but a larger magnetic
energy; \figref{Fig:35OC-ROs-jetvra}, bottom panel).  The
generation of the outflow even ceases occasionally and instead falls
back towards the core (see around $t \approx 700 \, \ms$ and
$t \approx 1 \, \sek$ at the north pole).  These interruptions of the
jet occur when the accretion stream in the course of its stochastic
change of location shifts from the equatorial regions to the pole and
squeezes the outflow.  We point to \figref{Fig:35OC-Rw-2d} for
$t=700\,\ms$ exemplifying the contrast between the low-entropy
downflow at the north and the high-entropy outflow at the south pole,
respectively.  At other times, the impact of the hydrodynamic
instabilities is less pronounced, but nevertheless notable
fluctuations of the velocity (\figref{Fig:35OC-Rw-jetvra}, bottom
subpanel) and of the energy flux at the jet base as well as higher radii
are present in this model. The latter quantity varies both on short
and long time scales by up to an order of magnitude.  As a
consequence, the explosion energy and mass do not grow as steadily as
in \modl{35OC-RO}, though they reach values similar to those of the
stronger magnetised, earlier exploding \modl{35OC-RO}
(\figref{Fig:35OC-glob1}).

The great variability of jet formation contributes to a rather complex
morphology of the ejecta (see $t = 0.7, 1.6 \, \sek$ in
\figref{Fig:35OC-Rw-2d}).  They consist of the two main components,
viz.\xspace beam and cocoon.  Both, however, are usually wider and
their cross sections change more along the axis than in more
magnetised models (compare Figs.\,\ref{Fig:35OC-Rw-2d} and
\ref{Fig:35OC-ROs-jets}).  Particularly strong modifications develop
at times when the injection is interrupted in one hemisphere as,
\eg~at $t =700 \, \ms$. As a result, the northern and southern polar
outflows are much more asymmetric than in case of
\modl{35OC-RO}. Whether this \emph{north/south}-asymmetry is
maintained until the ejecta breaks out of the surface of the stellar
progenitor is still uncertain (further episodes of interrupted
injection may develop after $1.6\,\sek$). However, the two polar
outflows are made out of unbound matter, which will probably emerge
asymmetrically from the stellar surface. Other models with a magnetic
field strength below that of the original stellar model
(e.g. \modl{35OC-Sw}) also display a qualitatively similar
north/south-asymmetry in the outflows, which is likely a distinctive
property of outflows generated from pre-supernova progenitors with
\emph{substellar} magnetisation.

\subsubsection{Magnetically driven explosions}
\label{sSek:ResDeMDE}

\begin{figure}
  \centering
  \includegraphics[width=\linewidth]{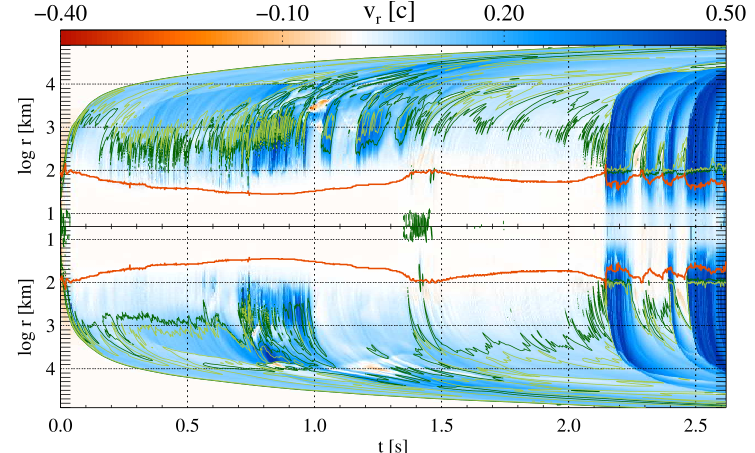}
  \includegraphics[width=\linewidth]{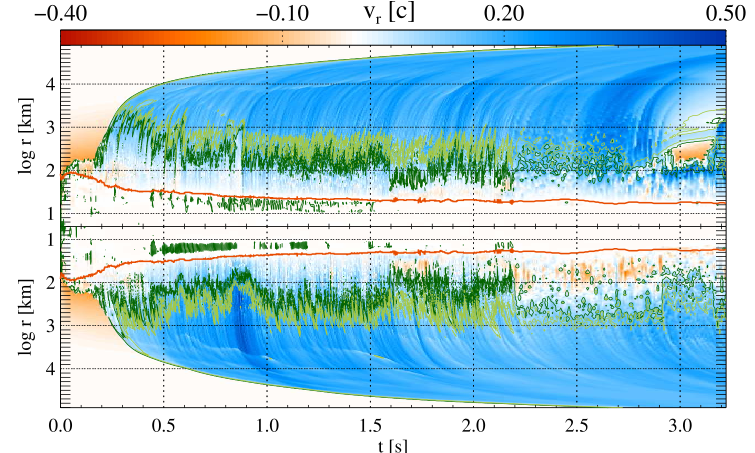}
  \caption{
    Same as \figref{Fig:35OC-Rw-jetvra} but for 
    \modls{35OC-Rs} (\banel{top}) and \modelname{35OC-RO}
    (\banel{bottom}).  
  }
  \label{Fig:35OC-ROs-jetvra}
\end{figure}

\begin{figure}
  \centering
  \includegraphics[width=\linewidth]{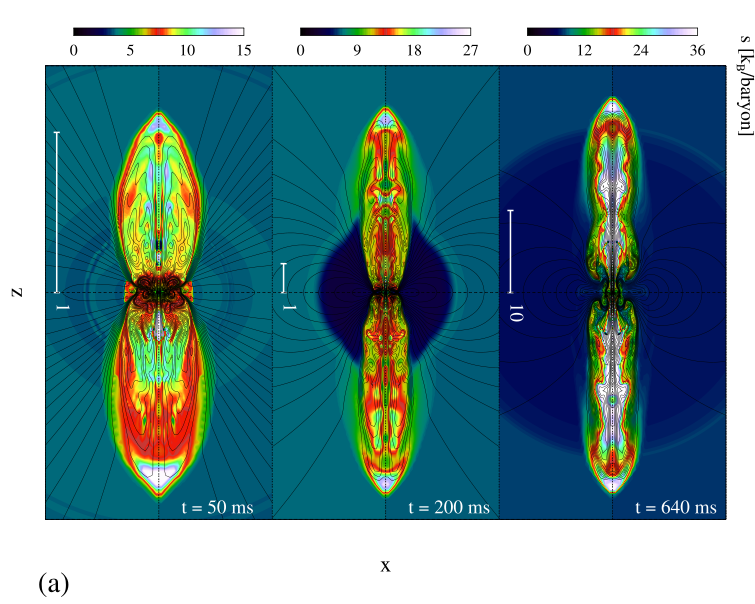}
  \includegraphics[width=\linewidth]{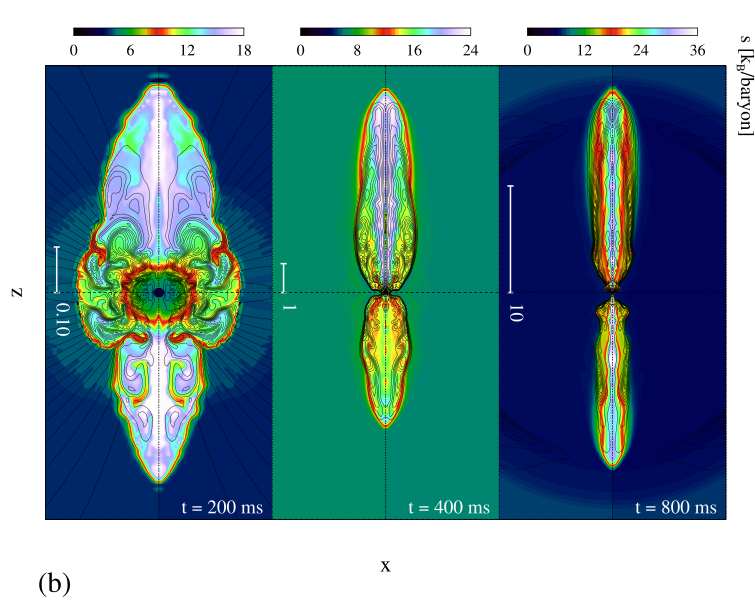}
  \caption{
    \banel{Top}: colour maps of the specific entropy and magnetic
    field lines in the explosion of \modl{35OC-Rs}.  The scales of the
    panels are indicated by rulers with a given length in units of
    $1000 \, \km$.
    \banel{Bottom}: same for \modl{35OC-RO}.
  }
  \label{Fig:35OC-ROs-jets}
\end{figure}

\begin{figure}
  \centering
  \includegraphics[width=\linewidth]{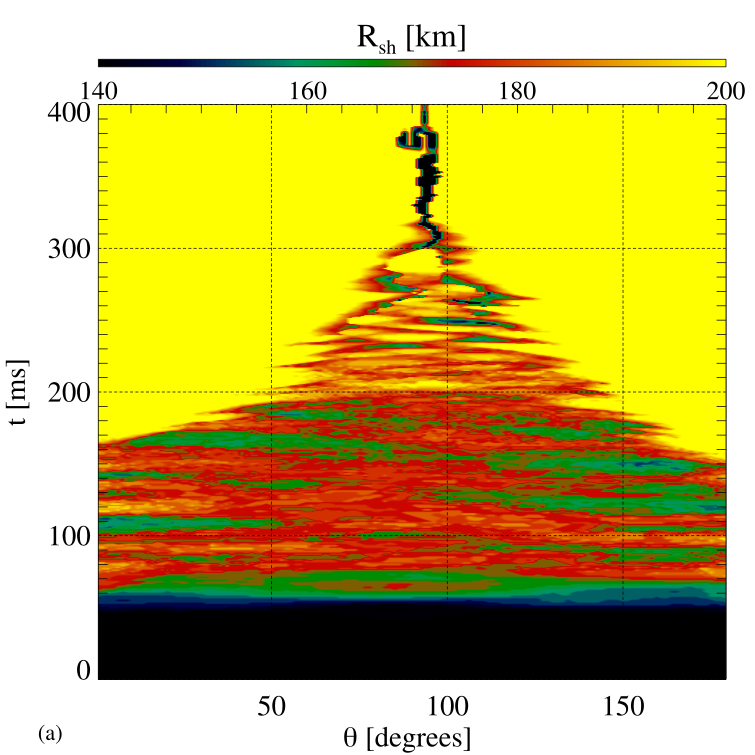}
  \caption{
    Same as \figref{Fig:35OC-Rw-shockrad}, but
    for \modl{35OC-RO}.
  }
  \label{Fig:35OC-RO-shockrad}
\end{figure}

The generation of such a column of strong, \ie super-equipartition,
magnetic field along the rotational axis in models like
\modelname{s20-3} and \modelname{35OC-Rs/RO/RO2/Rp2/3/4} leads to the
formation of polar outflows.  We show the structure of the ensuing
explosion for the least (\modelname{35OC-RO}) and most
(\modelname{35OC-Rs}) extreme of our models in
\figref{Fig:35OC-ROs-jets} and the evolution of the velocity along the
polar axis in \figref{Fig:35OC-ROs-jetvra}.

In both cases, the shock wave starts to expand along the axis long
before the weak-field version of the same progenitor, \modl{35OC-Rw},
achieves shock revival.  The prototypical case of \modl{35OC-Rs} even
launches the explosion without the shock wave ever stagnating along
the pole.  We note that the shock wave stagnates at lower
latitudes where the magnetic field is weaker.

Neutrino heating does not play an important role in \modl{35OC-Rs}.
Instead, the high positive radial velocities of the gas are driven by
a magnetic field that is locally in or above equipartition not only
with the flow, but also with the gas pressure.  The heads of the
outflows reach a radius of $r = 1000 \, \km$ at only $t = 50 \, \ms$
after bounce.  Already relatively early on after the explosion is
launched ($t = 0.2 \, \sek$), the outflow is narrowly collimated.  It
maintains a similar aspect ratio for the rest of the simulation
(\figref{Fig:35OC-ROs-jets}; $t = 0.64 \, \sek$).

The jets are dominated by the magnetic field whose energy exceeds both
internal and kinetic energies considerably.  The ratio between gas
pressure and magnetic pressure and the \Alfven number of the flow take
local values as low as $\beta = P / P_{\mathrm{mag}} \sim 0.01$ and
$\Alnum = |\vec v| / c_{\mathrm{A}} \sim 0.1$, respectively.  The
dominance of the magnetic field persists during the entire time we
simulated.  As the jet reaches higher radii, its interior can be
separated into two regions of sub-\Alfvenic and super-\Alfvenic speeds
inside and outside of a transition that fluctuates between radii of
several 100 and several 1000 km, respectively, for \modl{35OC-RO} (see
\figref{Fig:35OC-ROs-jetvra} showing where the outflows
are generated and how fluid elements propagate along the polar axis).
The (radial) outflow velocity (top panel of
\figref{Fig:35OC-ROs-jetvra}) varies significantly in the former
region following the dynamics of the field.  Only after passing into
the super-\Alfvenic region do fluid elements consistently show high
positive velocities.

Because the forces corresponding to the extremely strong fields act on
shorter time scales than the neutrino heating, the thermodynamical
state of the gas is not altered significantly by neutrinos.  As a
consequence, the early phase of the outflow is characterised by low
entropies and low electron fractions.  These properties have important
implications for the nucleosynthetic yields of the explosions,
similarly to the results of, \eg
\cite{Winteler_et_al__2012__apjl__MagnetorotationallyDrivenSupernovaeastheOriginofEarlyGalaxyr-processElements,Nishimura_et_al__2015__apj__Ther-processNucleosynthesisintheVariousJet-likeExplosionsofMagnetorotationalCore-collapseSupernovae,Nishimura_et_al__2017__apjl__TheIntermediater-processinCore-collapseSupernovaeDrivenbytheMagneto-rotationalInstability,Halevi_Moesta__2018__mnras__r-Processnucleosynthesisfromthree-dimensionaljet-drivencore-collapsesupernovaewithmagneticmisalignments}
finding r-process nucleosynthesis in magnetically driven outflows.  We
note that recent studies of the chemical evolution of the Galaxy
suggest that a substantial fraction of the r-process elements are
formed by sources other than binary mergers involving neutron stars,
with strongly magnetised classes of stellar core collapse being the
most likely additional site
\citep{Siegel__2019__arXive-prints__GW170817--thefirstobservedneutronstarmergeranditskilonova:implicationsfortheastrophysicalsiteofther-process,Cote_et_al__2019__apj__NeutronStarMergersMightNotBetheOnlySourceofr-processElementsintheMilkyWay}.

\Modl{35OC-RO} represents a transition between the class of MHD-driven
jets and that of neutrino-driven explosions.  The shock wave starts to
expand along the north and south poles at $t \approx 150 \, \ms$.
Shortly after the start of shock runaway ($t = 200 \, \ms$,
\banel{panel (b)} of \figref{Fig:35OC-ROs-jets}), we find radially
expanding high-entropy regions at high latitudes, whereas the
equatorial region is dominated by a roughly spherical gain layer in
which non-spherical instabilities produce bubbles of moderate extent
that appear, merge, and dissipate in a fast succession of events, but
that fail to generate a runaway of the shock.

\begin{figure}
  \centering
  \includegraphics[width=\linewidth]{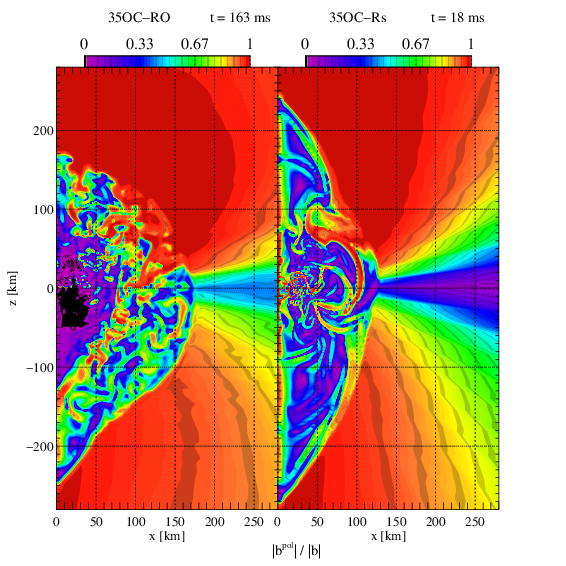}
  \caption{
    Ratio of the poloidal to the total magnetic field in
    \modls{35OC-RO} (left) and \modelname{35OC-Rs} (right) around the onset of the
    explosion (times shown at the top of the figure).
  }
  \label{Fig:2d-bpb}
\end{figure}

The success of polar, as opposed to equatorial, shock revival is
rooted in a combination of a column oriented along the polar axis in
which the magnetic field reaches equipartition with the kinetic
energy, and the pronounced anisotropy of the neutrino emission caused
by the rotational flattening of the PNS and, in particular, the
\nusps.  In a geometry similar to that found by
\cite{Burrows_etal__2007__ApJ__MHD-SN,Takiwaki_Kotake_Sato__2009__apj__Special_Relativistic_Simulations_of_Magnetically_Dominated_Jets_in_Collapsing_Massive_Stars,Takiwaki_Kotake__2011__apj__GravitationalWaveSignaturesofMagnetohydrodynamicallyDrivenCore-collapseSupernovaExplosions},
the poloidal and toroidal components of the magnetic field are roughly
equal over the largest part of the magnetic column.  In a narrow
region at its centre, however, the radial component dominates.  Its
importance is more pronounced for stronger initial fields, as we show
in \figref{Fig:2d-bpb} shows.  In the \modl{{35OC-RO}} (left), the
explosion starts from the south polar region at
$r \lesssim 200 \, \km$ where the poloidal field is fairly strong.
\Modl{35OC-Rs} (right part) exhibits a strong poloidal field at the
centre of the bipolar outflows in both hemispheres.  For the influence
of the field geometry, see the work by
\cite{Sawai_Kotake_Yamada__2005__ApJ__MHD-SN-nonuniform-field,Bugli_et_al__2019__arXive-prints__Theimpactofnon-dipolarmagneticfieldsincore-collapsesupernovae}.

In the phase leading up to the
onset of explosion, the deformation is moderate, and so is the
pole-to-equator difference of the neutrino fluxes.  Nevertheless, this
moderate degree of asymmetry (at $t = 150 \, \ms$, the neutrino flux
along the poles exceeds that at the equator by about $30 \, \%$) is
sufficient to focus the neutrino flux into cones around the poles
where it contributes to the local heating of the gas and to reverting
the infall.  
The
neutrino heating as well as the magnetic driving of the outflows are
affected indirectly by a reduced activity of hydrodynamic
instabilities compared to \modl{35OC-Rw} and, thus, less mixing
between different latitudes. The difference is caused by the stronger
magnetic field whose tension, resisting bending by the flow, partially
suppresses the non-radial flows.  This effect can be seen in the less
pronounced sloshing mode of the shock radius (compare
\figref{Fig:35OC-RO-shockrad} to \figref{Fig:35OC-Rw-shockrad}; shock
sloshing is characterized by zig-zagging patterns of red-to-yellow
shades in these figures).  The reduction can also be traced in a mean
$\theta$-component of the velocity in the gain layer that has only
about 2 thirds of the energy of \modl{35OC-Rw}.

\subsection{Interpretation}
\label{sSek:ResEC}

As we have discussed above, the competition between the timescales of
advection and neutrino heating based on the angular averages of the
stellar structure only provides a rough explanation of the success or
failure of the explosion.  We reiterate its most notable properties
and limitations:
\begin{itemize}
\item Most models show a correlation between shock revival and an
  increase of the ratio between advection and heating times.
\item The start of shock runaway and the exact moment at which
  $\tautauavg =1 $ may, however, be separated by a significant time.
  Quite commonly, the former precedes the latter, although it also may
  be the other way round such as in \modls{35OB-RO} and
  \modelname{35OB-RRw}, where both timescales are equal for about 200
  ms and almost 1 s before the explosions finally set in.
\item In several models, the rise of $\tautauavg$ is not caused by more
  efficient neutrino heating, but is a secondary effect of an
  expansion of the shock wave and, consequently, the rise of
  $\tauadv$, as an explosion is launched by magnetic
  stresses.
\item The angularly averaged values of the advection and heating times
  cannot account for the very asymmetric geometry of the fastest
  rotators, causing large differences between the polar and equatorial
  neutrino emission.
\end{itemize}

In the following, we will explore several modifications and
alternatives to this criterion in order to account for the effects of
the deformation of the core by rotation and for the magnetic fields.

\subsubsection{Angle-dependent analysis}

\begin{figure*}
  \centering
  \includegraphics[width=\linewidth]{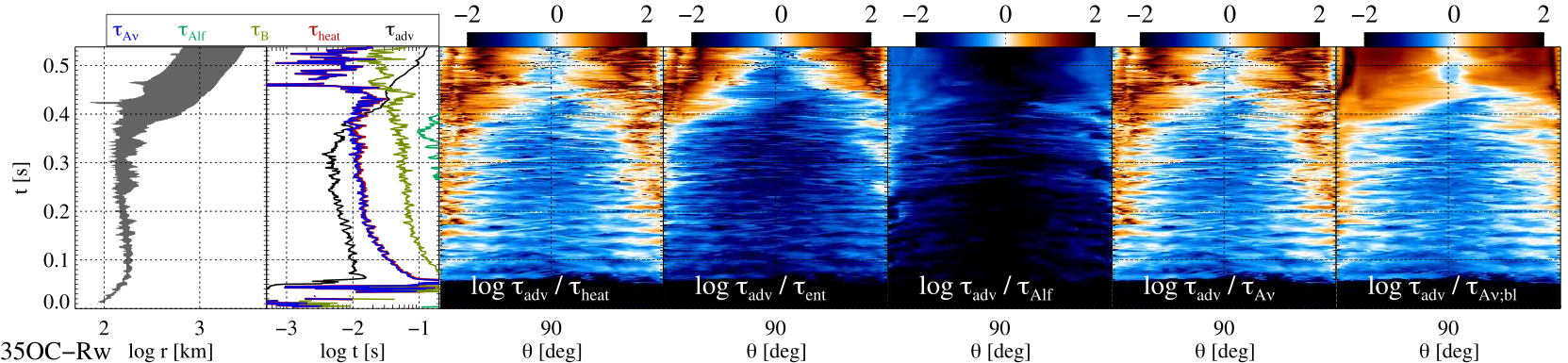}
  \includegraphics[width=\linewidth]{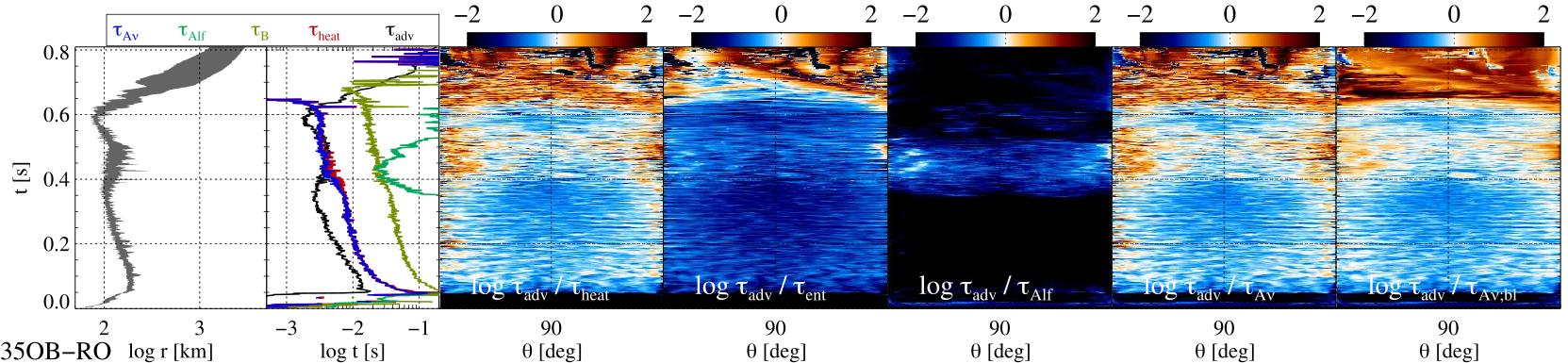}
  \includegraphics[width=\linewidth]{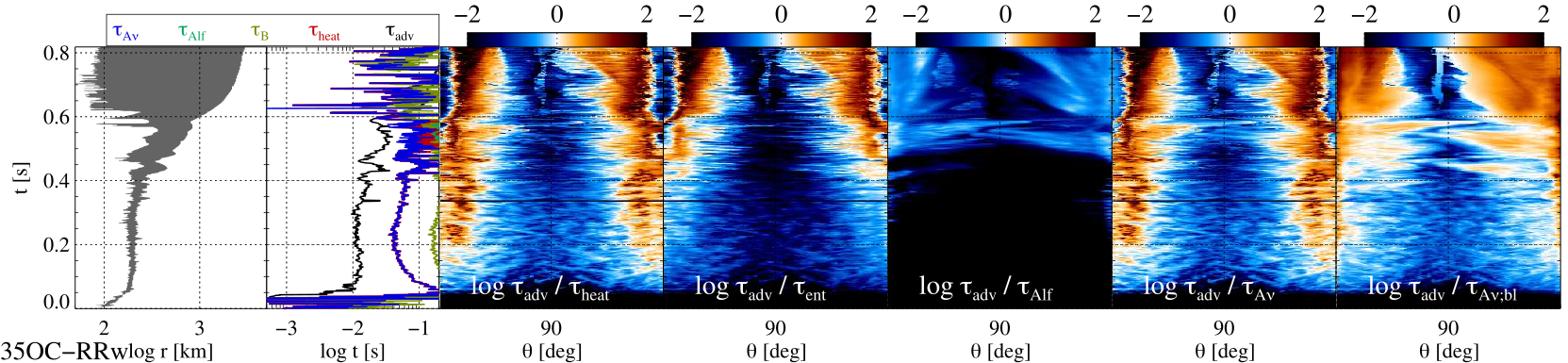}
  \includegraphics[width=\linewidth]{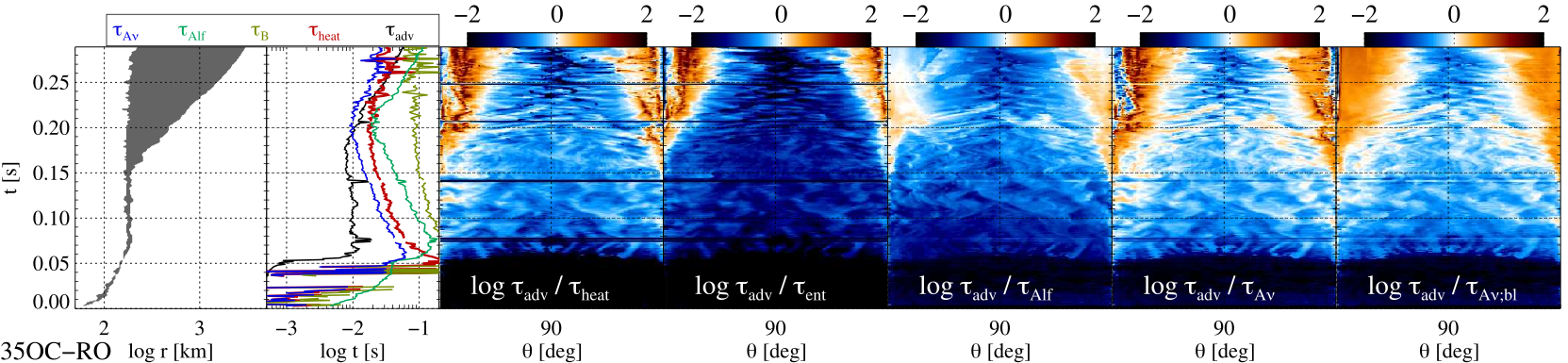}
  \includegraphics[width=\linewidth]{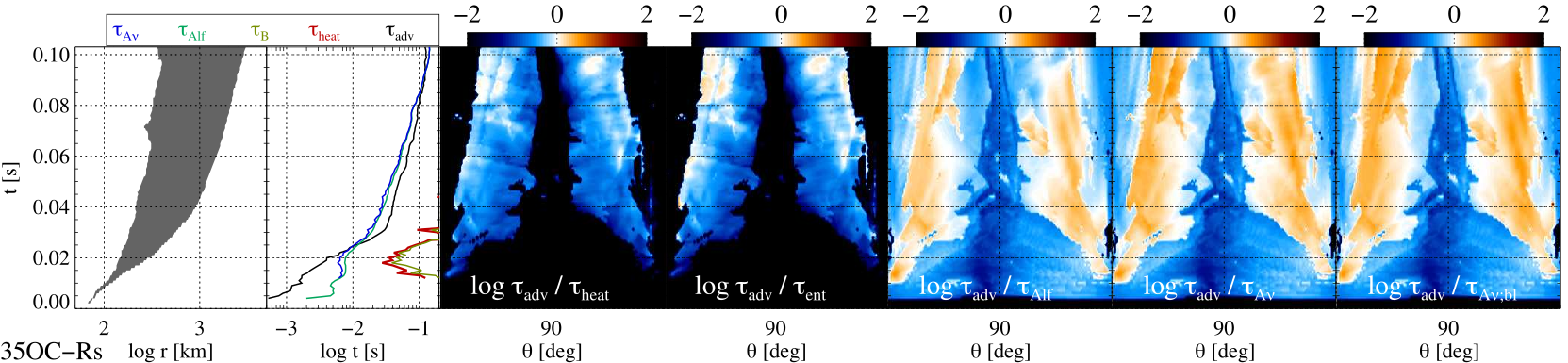}
  \caption{
    Comparison of the shock location and the time scales
    introduced in \secref{sSek:ResEC} for various models, as indicated
    in the row.  The left subpanel of each row shows the time evolution of the shock radii (grey
    band delimited by the minimum and maximum shock radii).  The
    second subpanel presents the time evolution of the advection,
    heating, enthalpy, \Alfven, and combined timescales (distinguished
    by colours as shown at the top).  The following  four subpanels show the
    timescale ratios as a function of time and 
    the latitude (left to right:
    $\tauadv / \tauhtg$, $\tauadv / \tauent$,  $\tauadv / \tauAlf$,
    $\tauadv / \tauAnu$).  The right panel displays $\tauadv /
    \tauAnubl$ including the effect of angular and temporal
    mixing.
  }
  \label{Fig:Excrits-alltaus}
\end{figure*}

Instead of angular integrals/averages of the relevant quantities, we
perform an analysis of the advection and heating times on each
$\theta$-angle separately.  We compute the local advection and heating
timescales, $\tau_{\textrm{adv; heat}}(\theta)$, by replacing the
angular averages in the definitions of the variables by the
corresponding angle-dependent quantities, \eg:
\begin{eqnarray}
  \label{Gl:tadvtheta}
  \tauadv (\theta) & = &
                                     \frac{D ( \theta )} {|\langle v^r \rangle (\theta) |},
  \\
  \label{Gl:thtgtheta} 
  \tauhtg ( \theta) & = &
                                       \frac{- \Egain(\theta) } {  Q_{\nu} (\theta)}.
\end{eqnarray}
The radial extent of the gain layer, $D (\theta)$, the radial average
of the radial velocity over the gain layer at a given angle $\theta$,
$|\langle v^{r}\rangle (\theta) |$, and the radial integrals of the
total energy and the neutrino heating, $\Egain (\theta)$
and $Q_{\nu} (\theta)$, are necessarily more noisy than their integral
versions.  Nevertheless, they allow for several important
observations, for which we refer to the subpanels marked as
  $\log \tautau$ in \figref{Fig:Excrits-alltaus}.
For 
several models, we display, from left to right, the time evolution of
the shock radius (the grey shades denoting the minimum and maximum
shock radii), the integral advection, heating, and \Alfven timescales
(the latter will be introduced below), and, as functions of
time and latitude, the logarithm of $\tautauth$, of the ratio between
advection and \Alfven timescales, and of the Bernoulli parameter (see
below).

The angle-dependent analysis reveals that the conditions most
favourable for explosions are developing at the axes (note the reddish
regions at $\theta = 0, 180 \grad$ in \figref{Fig:Excrits-alltaus}).
The enhancement of $\tautauth$ \wrt the equator and the angular
averages is most notable for rapidly rotating models.  We point in
particular to \modl{35OC-Rw} (top row).  Already about
200 ms before the shock starts its rapid runaway, neutrino heating is
faster than mass accretion along both axes, while lower latitudes show
the opposite behaviour.  As described above, the pole-to-equator
contrast is a consequence of the concentration of the neutrino
emission towards high latitudes caused by the strong deformation of
the PNS.  During this phase, the shock wave does not recede, but shows
pronounced oscillations, which are largest at the pole.  Hence, the
local analysis seems to predict a much earlier explosion, while the
equality between the angularly averaged timescales (the intersection
of the black and red lines in the second subpanel) lags the onset of
explosion.  An appropriate description of the explosion mechanism
should therefore combine local and global elements.

The time of explosion of \modl{35OB-RO} coincides well with the
transition of both the global, $\tautauavg$, and the local,
$\tautauth$, ratio of timescales to values exceeding unity, though
polar regions of the model exhibit $\tautauth > 1$ already quite early
without an explosion setting in.  For the faster rotating
\modl{35OC-RRw}, on the other hand, the discrepancy between the local
and global timescale ratios and the start of the explosion are more
pronounced due to the higher concentration of the neutrino emission
towards the poles.  Equality between the global advection and heating
time scales is achieved after the shock wave has already reached a
maximum radius of around 1000 km, while the polar regions exhibit
favourable conditions for shock revival long before that.

Both versions of the timescale criterion fail in the strongest
magnetised models (see panels for \modl{35OC-Rs} and, to a lesser
degree, \modelname{35OC-RO}).  Their explosions develop without
$\tautau (\theta)$ exceeding unity even at the poles. The timescale
ratio increases beyond unity only afterwards as the advection time
increases during the shock expansion.

\subsubsection{Specific enthalpy}

At its core, the criterion based on the timescales of advection and
neutrino heating is based on an estimate of the total energy: an
explosion is likely if the post-shock matter gains energy sufficiently
fast to reach positive total energy, \ie to become gravitationally
unbound.  In its global version, the control volume in which the gas
has to adquire positive energy is the entire gain layer, whereas the
local version checks under individual angles in a ray-by-ray manner.
Many of our models, however, indicate that the positivity of the total
energy is not sufficient for shock revival.  We find unbound regions
of sometimes quite large extent in the gain layers of several models
without the accretion shock becoming unstable and turning into an
explosion.  These bubbles may persist for a long time and remain at
roughly the same location, usually at high latitudes, albeit with
changing sizes and geometries.  

The requirement on the positive total energy comes short in one
important aspect: it does not account for the ram pressure of the
infalling matter.  For the shock wave to be reverted, the post-shock
gas not only has to achieve positive energy, but it has to be capable
of reverting the infall of the outer layers.  We thus turn to another,
related version of describing the explosion conditions.  We first note
that a fluid element falling adiabatically in a gravitational field
conserves the Bernoulli parameter or total specific enthalpy,
\begin{equation}
  \label{Gl:Bernoulli}
  h_{\mathrm{tot}} (\theta) = 
  ( e_{\textsc{mhd}}  (\theta) + P_{\star} (\theta) ) / 
  \rho  (\theta) + \phi  (\theta).
\end{equation}
If the gas is subject to an external heating source such as neutrinos,
the internal energy increases at a rate $\dot{e}_{\mathrm{int}} =
q_{\nu}$, where $q_{\nu}=\alpha Q_\star^0/\rho$.  For an ideal gas
with an adiabatic index $\Gamma$, this change corresponds to an
increase of the gas pressure at a rate $\dot{P} = (\Gamma - 1 )
\dot{e}_{\mathrm{int}}$ and, thus, $h_{\mathrm{tot}}$ increases at a
rate $\dot{h}_{\mathrm{tot}} = \Gamma q_{\nu}$.  In this view, an
explosion might develop if the fluid element increases its
$h_{\mathrm{tot}}$ while falling through the gain layer such as to
exceed the specific ram pressure of the pre-shock gas,
$p_{\mathrm{ram}}/\rho = (v_{\mathrm{ps}}^r)^2$.  
For a better comparison to the discussion above, we introduce a new
timescale, $\tauent$, as the time it takes for heating to increase the
total enthalpy in the gain layer to a value corresponding to the
pre-shock ram pressure.  In its angle-dependent form, it is implicitly
given by:
\begin{equation}
  \label{Gl:tauent}
    h_{\mathrm{tot}} (\theta) 
    + \tauent \dot{h}_{\mathrm{tot}} (\theta)
    =  ( p_{\mathrm{ram}}/\rho ) (\theta).
\end{equation}
We surmise that $\tauent < \tauadv$ is favourable for shock revival,
and, hereafter, we shall say that a model fulfills the Bernouilli or
enthalpy criterion when $ \tauadv/\tauent >1$.

We show the ratio $\tautauentth$ in the second of the time-latitude
subpanels of \figref{Fig:Excrits-alltaus}.  Compared to the original
timescale analysis, the space-time regions where the Bernoulli
criterion is fulfilled are considerably smaller.  Hence, this
modification is less prone to overestimate the tendency of the gain
layer to produce shock revival.  To show this, we refer in particular
to the case of \modls{35OB-RO}, \modelname{35OC-Rw}, and
\modelname{35OC-RRw} which exhibit smaller regions indicating shock
revival when applying the Bernoulli criterion than for the timescale
comparison.  We note that models exploding predominantly by magnetic
stresses again differ from neutrino-driven explosion by requiring only
marginal fulfilment of the Bernoulli criterion, as the white rather
than red regions at the poles of \modls{35OC-RO}, and
\modelname{35OC-Rs} demonstrate.

\subsubsection{Timescales including the magnetic field}

The failure of the local and global analysis of $\tautau$ to explain
the explosions of the models with strongest magnetic fields is a
direct consequence of the predominantly magnetic nature of these
explosions.  We could connect them to the appearance of very strong
magnetic fields along the polar axis where the flow has sub-\Alfvenic
speeds.  This connection suggests to introduce another timescale, that
associated to the propagation of \Alfven waves through the gain layer.
Hence, we define the \Alfven timescale as
\begin{equation}
  \label{Gl:tauAlf}
  \tauAlf ( \theta  ) = \frac{D (\theta )} { | \langle c_{A} \rangle|( \theta )},
\end{equation}
where $c_{A}=|B|/\sqrt{\rho}$ is the \Alfven speed and
$\langle . \rangle$ indicates a radial average over the gain region at
angle $\theta$.  We propose that a magnetically driven explosion can
be triggered if the ratio between the advection and the \Alfven times
exceeds unity.  Such a situation would correspond to energetic
equipartition between the magnetic field and the velocity.  Hence the
magnetic pressure would be strong enough to counteract the infall of
the gas and turn it into an explosion.  We note that we compute the
\Alfven timescale from the total magnetic field.  An alternative
definition might be based on only, \eg, the poloidal or toroidal
components.  However, as we discuss in \secref{sSek:ResDeMDE}, the two
components tend to be of the same order in the regions where the
explosion is initiated.  Hence, such a timescale based on one
component only would not affect the outcome of the analysis.

Our results support this proposition: \modls{35OC-RO}
and \modelname{35OC-Rs} exhibit $\tautaumth > 1$ at the poles when the
shock starts its rapid expansion, as we show in third time-latitude
subpanels of \figref{Fig:Excrits-alltaus} where $\tautaumth > 1$,
corresponding to red colours, at $\theta = 0, 180 \grad$ when the
explosion sets in.  We note that a global version of this criterion
based on the angularly averaged values, $\tautaumavg$, does not
produce reasonable results for these rather collimated explosions in
which the rapid launch at the poles is largely disconnected from the
dynamics at lower latitudes.

We compute a combined timescale from the magnetic and the heating
times,
\begin{equation}
    \label{Gl:tauAnu}
    \tauAnu (\theta) = 
    ( \tauAlf^{-1} (\theta) + \tauhtg^{-1}(\theta))^{-1},
\end{equation}
and show the result in the fourth time-latitude subpanels of
\figref{Fig:Excrits-alltaus}.  By construction, if $\tauAlf$ and
$\tauhtg$ differ by a large factor, $\tauAnu$ is equal to the lesser
of the two.  Hence, it agrees with $\tauhtg$ in \modls{35OC-Rw},
\modelname{35OB-RO}, and \modelname{35OC-RRw} and with $\tauAnu$ for
\modl{35OC-Rs}.  In \modl{35OC-RO}, on the other hand, $\tauAnu$ is
significantly less than either of the other two.  Consequently, it
equals the advection timescale earlier than those, leading to a better
agreement with the onset of shock expansion.  This improvement is most
notable for the angle-dependent quantity, whereas the angularly
averaged version achieves equality with $\tauadv$ after the shock
starts to expand along the poles.  Thus, the criterion
$\tauadv/\tauAnu>1$, though not completely perfect, is the best among
the global criteria to describe when favourable explosion conditions
develop in the gain layer.

\subsubsection{Mixing}

As we have seen above, some models are adequately described by the
angle-dependent timescale analysis, while others are better described
by global quantities.  In general, however, the best description seems
to be a mixture between the global and the angle-dependent one as
locally at the poles $\tautauAnu$ can exceed unity for a long time
without leading to an explosion.  Such a behaviour is more common
among models whose explosions are driven by neutrinos rather than by
magnetic fields; among the former group, it particularly affects
rapidly rotating models.

We attribute the possible mismatch between the local timescale ratio
and the actual onset of the explosion to the intensity of lateral
mixing throughout the gain layer.  If fluid elements are exchanged
between the polar regions where heating is most effective and the
equatorial ones where it is less intense, the heating efficiency may
be reduced significantly compared to the case where a fluid element
entering the gain layer at the pole always stays there and is, hence,
always exposed to strong heating.  Essentially, the heating received
at the poles is diluted across a wider range of angles.

We propose to account for the mixing by blurring the neutrino heating
timescale across a domain in the angular direction whose extent
depends on the lateral velocity and across the dwell time of a fluid
element in the gain layer, $\tauadv$.  During that time, the fluid
element will on average be displaced in $\theta$-direction by a
distance which we estimate as
$d_{\theta} (\theta) \approx \langle v_{\theta} (\theta) \rangle
\tauadv$, where we average the $\theta$-component over the gain layer
at a fixed angle,
$ \langle v_{\theta} (\theta) \rangle= \left[ \int \mathrm{d}r\, r^2 (
  v^{\theta} (r,\theta) )^2 / \int \mathrm{d}r\, r^2 \right]^{1/2}$,
where the integrals are to be taken over the gain layer.  As a
consequence, its effective neutrino heating rate is an average of
those inside this domain.  Given $\tauadv ( \theta, t)$ and
$ \langle v_{\theta} (\theta,t) \rangle$ at time $t$ and angle
$\theta$, we compute averages of the heating timescale over
$t \pm \tauadv$ and
$\theta \pm \langle v_{\theta} (\theta) \rangle \tauadv /
R_{\mathrm{gain}}$, where $R_{\mathrm{gain}}$ is the mean radius of
the gain layer.  We will refer to the blurred version of the heating
timescale as $\tauhtgbl$.  We combine this quantity with $\tauAlf$ to
define the timescale $\tauAnubl$ in the same way as in
\eqref{Gl:tauAnu}, \ie,
\begin{equation}
  \label{eq:tauAnubl}
  \tauAnubl (\theta) = ( \tauAlf^{-1} (\theta) + \tauhtgbl^{-1}(\theta))^{-1}.
\end{equation}
Note that we do not apply the same blurring
procedure to the \Alfven timescale because we find that the magnetic
forces launching an MHD explosion act mostly along the (radial) field
lines without spreading across a wider range of latitudes.  The right
time-latitude subpanels of \figref{Fig:Excrits-alltaus} presents the
resulting ratio $\tautauAnubl$.

Neutrino heating is most important for \modls{35OC-Rw},
\modelname{35OC-RRw}, \modelname{35OB-RO} and to a certain degree also
\modelname{35OC-RO}.  Mixing reduces the maxima of $\tautauAnu$ along
the poles. This effect is most notable in the reduced extent of the
reddish regions in \modls{35OC-Rw} (before $t \approx 350 \, \ms$) and
\modelname{35OC-RRw} (before $t \approx 400 \, \ms$). The reduction,
underscoring that very localised heating may be ineffective at
reviving the shock, improves the consistency between the (local) ratio
of timescales and the evolution of the shock radii.

Stronger magnetic stresses partially suppress lateral motions and
hence mixing.  Thus, mixing leads to only minor changes to the ratio
of timescales.  The strongest magnetised models such as \modl{35OC-Rs}
are less affected by lateral mixing.  The blurred timescale ratio
gives essentially the same results as the original one.
\Modl{35OC-RO} gain represents an intermediate case in which mixing is
less important than in the models with weaker magnetic fields.  We
therefore conclude that the criterion $\tautauAnubl$ is the best
performing on all the local criteria, underpinning the relevance of
lateral mixing in the gain layer.

\subsection{Outlook: three-dimensional  models}
\label{sSek:3d}

While performing and analyzing the simulations presented in this
article, we began to simulate a limited number of models in
unrestricted three-dimensional geometry in order to explore how much
the condition of axisymmetry distorts the evolution of the models.
The physics of the simulations is the same as in the axisymmetric
versions, while computing time restrictions allowed only for a lower
grid resolution of $64 \times 128$ in $\theta$ and $\phi$-directions,
respectively.  We note that we defer a full analysis of the models to
a later instant after having completed more of them and also after the
influence of the reduced grid resolution has been assessed.  For now,
we only present a brief summary of the evolution of models
\modelname{35OC-RO} and \modelname{35OC-Rs}, which run well into
the explosion.

\begin{figure}
  \centering
  \includegraphics[width=0.98\linewidth]{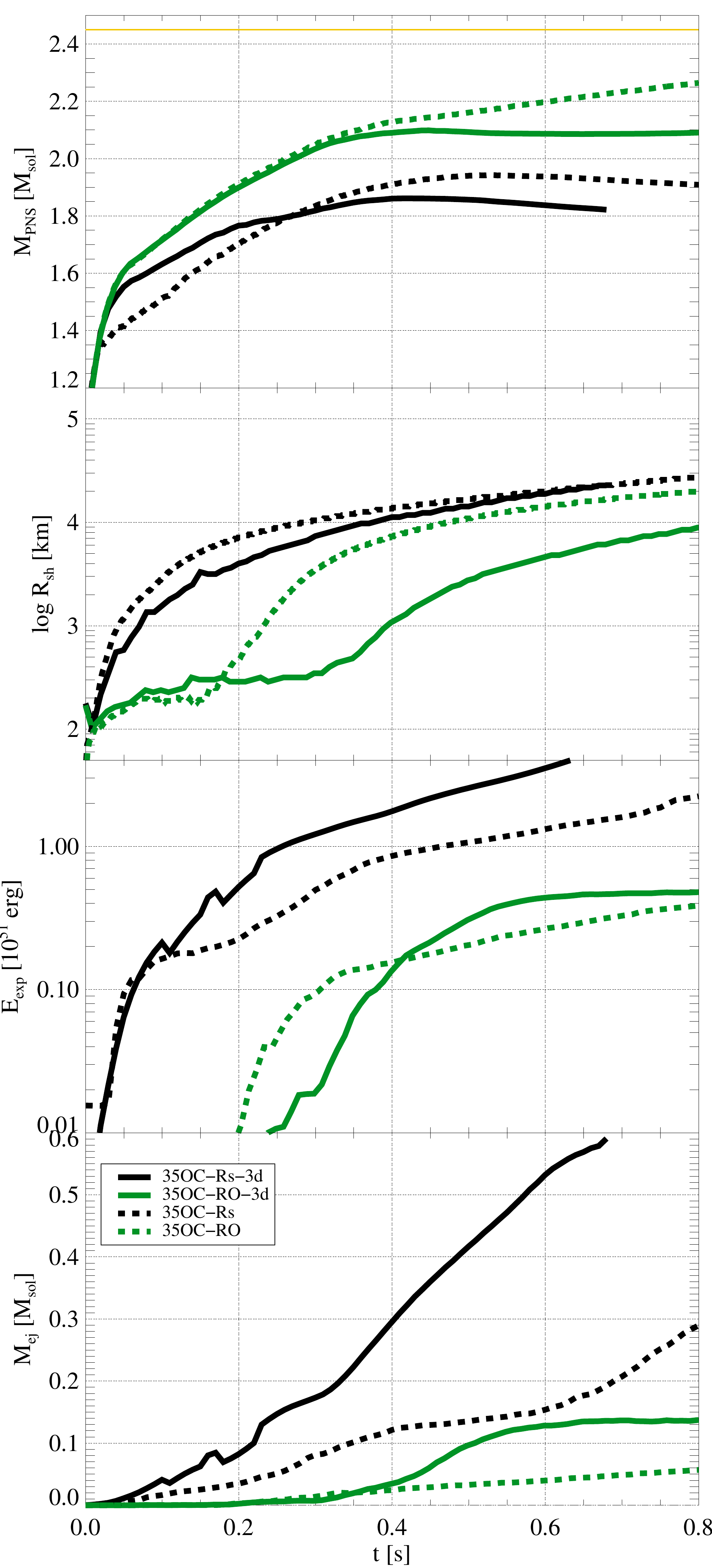}
  \caption{
    Evolution of global quantities characterising the 3d models.  From
    top to bottom, we show %
    the mass of the PNS, $M_\pnss$, %
    the maximum shock radius, $R_{\mathrm{sh; max}}$,
    the mass of the unbound ejecta, $M_{\mathrm{ej}}$, 
    and %
    the diagnostic explosion energy, $E_{\mathrm{exp}}$. 
    Line colours
    distinguishing the two models, \modelname{35OC-Rs/O}, as indicated
    in the bottom panel.  In comparison, the corresponding axisymmetric
    models are represented by dashed lines.
  }
  \label{Fig:3dglob}
\end{figure}

\begin{figure}
  \centering
  \includegraphics[width=\linewidth]{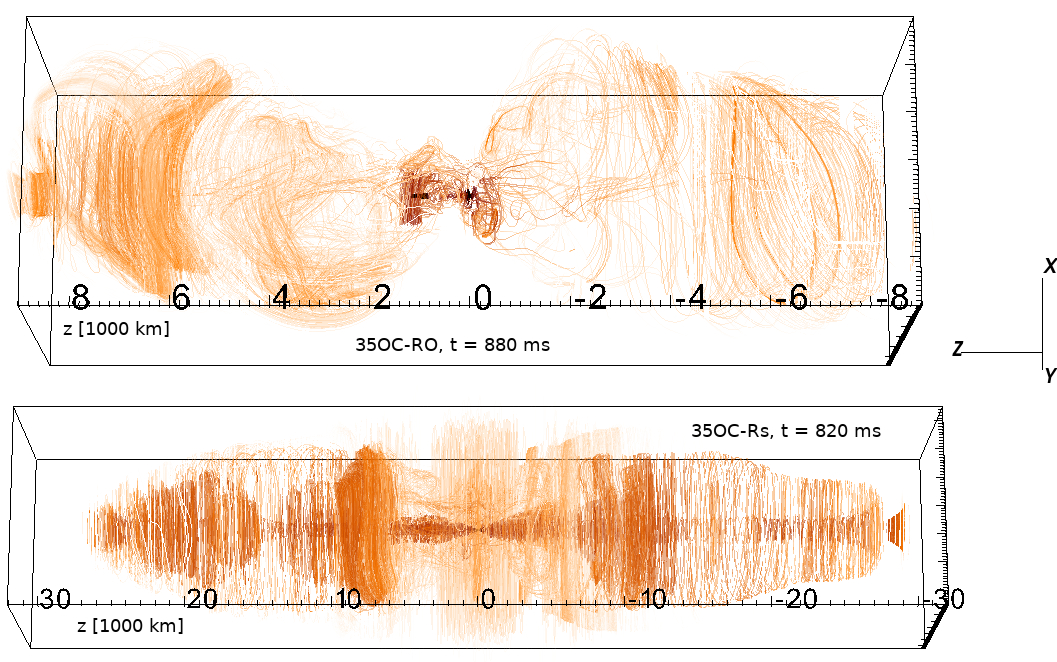}
  \caption{
    Structure of the three-dimensional versions of \modls{35OC-RO}
    (top)
    and \modelname{35OC-Rs} (bottom) at more than 800 ms after
    bounce.  We show field lines, with the colour indicating the field strength.
  }
  \label{Fig:3dmodels}
\end{figure}

We present the evolution of global properties of the models in
\figref{Fig:3dglob} and show their structure in \figref{Fig:3dmodels}.
Most importantly, both simulations produce explosions qualitatively
similar to the axisymmetric models.  There is, however, a trend to
explode slightly later in 3D than in 2D models.  \Modl{35OC-Rs}
explodes promptly after bounce, driven by the magnetorotational
stresses (see the $R_{\rm sh}$ panel of \figref{Fig:3dglob}).  The
delay until shock revival of \modl{35OC-RO} is about 300\,ms, \ie
$\sim 120\,$ms later than the axisymmetric version of the model
(\tabref{Tab:Exprops}).  Like in 2d, the mass of the unbound ejecta
and their (diagnostic) explosion energy increase very quickly for
\modl{35OC-Rs}.  The strong explosion very rapidly inhibits further
accretion onto the PNS, causing its mass to stay below
$M_\pnss \approx 2 \, \msol$ (\figref{Fig:3dglob}).
\Modl{35OC-RO} explodes less violently in 3D.  Both ejecta mass and
explosion energy seem to level off already before 0.6\,s, though a
longer simulation time would be required to test this statement.  The
PNS achieves a higher mass of $M_\pnss \approx 2.25 \, \msol$
($M_\pnss \approx 1.95 \, \msol$) in the axisymmetric version of
\modl{35OC-RO-3d} (\modelname{35OC-Rs-3d}).  For \modl{35OC-RO-3d}, if
the flat trend at late times were to be extrapolated, a collapse to a
BH seems unlikely in this case, compared to the late evolution
observed in \modl{35OC-RO}. We observe that there is a subtle trend to
increase the PNS mass starting for $t\gtrsim 0.7\,$s in
\modl{35OC-RO-3d}, which renders the previous extrapolation a bit
uncertain.  Furthermore, the growth of the PNS mass may display a
  non-monotonic behavior, depending on a complex interplay between
  accretion and magnetic field growth in the PNS and in the
  surrounding layers, as we have observed in many axisymmetric
  variants of the stellar core 35OC (see,
  e.g. \figref{Fig:35OC-glob1}). Thus, on the very basis of the
  computed evolution we cannot disregard that a BH forms also for
  \modl{35OC-RO-3d}.  Note that the qualitative existence of a global
maximum in the PNS is independent of the dimensionality of the model,
though the maximum mass attained by \modl{35OC-Rs-3d} is
$\sim 0.15\msol$ smaller than that of \modl{35OC-Rs}. Anyway, in none
of the two cases (\modelname{35OC-Rs} and
\modelname{35OC-Rs-3d}) the formation of a BH is likely.

In both 3D models, the explosion geometry is characterised by bipolar
outflows directed along the rotational axis
(\figref{Fig:3dmodels}).  Magnetic fields are the most
important (\modelname{35OC-RO-3d}) or the sole
(\modelname{35OC-Rs-3d}) driving force of the explosion.  While
non-axisymmetric structures develop and lead to, \eg spiral patterns
in the accretion onto the PNS, we do not observe a disruption of the
outflows by kink modes as found by
\cite{Mosta_et_al__2014__apjl__MagnetorotationalCore-collapseSupernovaeinThreeDimensions}.
Hence, we tentatively suggest that our axisymmetric models reasonably
approximate the dynamics of the cores at least at a qualitative level.

\section{Summary and conclusions}
\label{Sek:SumCon}

We followed the evolution of the cores of several rotating and
magnetised stars of zero-age-main-sequence masses $M_{\mathrm{ZAMS}} =
20$ and $ 35 \, \msol$ of different metallicities
after the onset of collapse to a PNS in multi-dimensional simulations.
Our main goal was to study potential CCSN explosions at masses that
are significantly higher than the more commonly investigated range of
masses.  Those of initial models with $M_{\mathrm{ZAMS}} = 35 \,
\msol$ are, like many stars in this regime, characterized by a fairly
compact density profile, for which they can be expected to lie in the
transition between stars likely to revive the stalled supernova shock
wave and stars whose high rates of mass accretion onto the shock wave
prevent an explosion
\citep{OConnor_Ott__2011__apj__BlackHoleFormationinFailingCore-CollapseSupernovae}.
Such a marginal situation might, in principle, allow for various
different evolutionary paths depending on variations in the
pre-collapse cores.  Therefore, we aimed at exploring this range of
possible outcomes by comparing three pre-collapse models taken from
stellar evolution calculations and, within each model, varying the
rotational profile and magnetic field.  We note that the stellar
evolution models in two of our initial models, \modelname{35OC} and
\modelname{35OB} of
\cite{Woosley_Heger__2006__apj__TheProgenitorStarsofGamma-RayBursts},
explicitly include rotation and magnetic fields, albeit in a spherical
approximation, whereas we artificially added angular momentum and
magnetic fields to the third core, \modelname{z35} of
\cite{Woosley_Heger_Weaver__2002__ReviewsofModernPhysics__The_evolution_and_explosion_of_massive_stars}.

We approached these issues by performing state-of-the-art axisymmetric
simulations coupling special-relativistic MHD with a
neutrino-transport scheme based on the two-moment formulation of the
spectral transport equation (see Sect.\,\ref{Sek:PhysNum}).

We simulated 19 models in axial symmetry: five versions of core
\modelname{s20}, ten of core \modelname{35OC}, and two each of cores
\modelname{35OB} and \modelname{z35}.  In \modls{35OC} and
\modelname{35OB}, most simulations used the original rotational
profile of the stellar-evolution calculations, but a few control
models were run with decreased and increased angular velocities.  Some
of the simulations of each core were run with the original magnetic
field, others with an artificial magnetic field of mixed
poloidal-toroidal topology and different normalization.  The
simulations of cores \modelname{s20} and \modelname{z35}, which
contain neither rotation nor magnetic fields, use a rotational profile
inspired by that of \modelname{35OB} and artificial magnetic fields.
Simulations were run until the cores collapsed to a BH
or, if failing to do so, for long times of several seconds.

Our main results can be summarized as follows:
\begin{itemize}
\item Despite the high compactness of all of the cores, most of our
  simulations yield successful explosions.  In most cases, the shock
  revival occurs within a fairly short time (around half a second)
  after bounce.  The explosion occurs when the PNS has accreted the
  entire inner core up to an interface where the density jumps by a
  factor of a few and, hence, the ram pressure of the infalling gas
  decreases.  The location of the interface depends on the specific
  core; for \modls{35OC} and \modelname{35OB}, it is a mass coordinate
  of $M_{\mathrm{IF}} \gtrsim 2.2 \msol$ and
  $M_{\mathrm{IF}} \gtrsim 2.3 \msol$, respectively.  Consequently,
  the PNS has a mass fairly close to the maximum mass allowed by the
  EOS already at the time when the explosion sets in.
\item Except for the strongest magnetic fields, the stalled shock is
  revived chiefly by neutrino heating.  
\item All explosions possess an asymmetric, prolate geometry.  This
  tendency is most pronounced for \modl{35OC-Rs}, combining the rapid
  rotation of star \modl{35OC} with a strong dipolar magnetic field,
  which develops a collimated, mildly relativistic magnetorotational
  explosion immediately after bounce.
\item However, magnetic fields are not necessary for an asymmetric
  explosion as all rapid rotators that explode at all do so in a
  prolate way.  In their case, the asymmetry is caused by a large
  pole-to-equator difference of the neutrino heating.  Centrifugal
  forces lead to a strong flattening of the PNS that focuses the
  neutrino emission into the polar direction, where consequently the
  conditions for shock revival are most favourable.
\item Besides causing a predominantly polar neutrino emission, rapid
  rotation tends to reduce the total neutrino luminosity because the
  centrifugal force halts mass accretion at a higher radius and, thus,
  less gravitational energy is released.  In the most extreme cases,
  the reduction is sufficient to prevent an explosion.  We note that
  this is not the case for any of the models computed with the
  original rotational profile obtained from stellar evolution.  Hence,
  whether any realistic core undergoes such an evolution remains to be
  seen.
\item Our models comprise a broad range of explosion energies up to
  several $\zehn{51} \, \erg$ for \modl{35OC-Rs} and ejecta masses of
  up to half a solar mass.  In many of them, these values have not yet
  reached saturation at the end of the simulation. Ongoing
    prolongations of some of the models presented here, show explosion
    energies $\sim \zehn{52}\,\erg$, broadly compatible with the
    energy released by hydrogen-poor superluminous supernovae
    \citep[e.g.][]{Kasen_Bildsten__2010__apj__SupernovaLightCurvesPoweredbyYoungMagnetars,Woosley__2010__apjl__Bright_SNe_from_Magnetar_Birth,Chatzopoulos_et_al__2013__apj__AnalyticalLightCurveModelsofSuperluminousSupernovae:ensuremathchi2-minimizationofParameterFits,Nicholl_et_al__2013__Nature__Slowlyfadingsuper-luminoussupernovaethatarenotpair-instabilityexplosions,Greiner_et_al__2015__Nature__Averyluminousmagnetar-poweredsupernovaassociatedwithanultra-longgamma-rayburst}
    or hypernovae \citep[e.g.][]{Iwamoto_et_al__1998__nat__Ahypernovamodelforthesupernovaassociatedwiththeensuremathgamma-rayburstof25April1998,Soderberg_et_al__2006__nat__RelativisticejectafromX-rayflashXRF060218andtherateofcosmicexplosions}.
\item We propose several ways to characterise the explosion mechanisms
  at work in our models.  The explosion criterion based on the balance
  between the advection and the neutrino-heating timescales
  \citep[see, \eg ][]{Janka__2001__aap__Conditionsforshockrevivalbyneutrinoheatingincore-collapsesupernovae,Thompson_Quataert_Burrows__2004__ApJ__Vis_Rot_SN,Murphy_Burrows__2008__apj__CriteriaforCore-CollapseSupernovaExplosionsbytheNeutrinoMechanism}
  in the gain layer yields an approximate agreement with the onset of
  explosion for many models, when evaluated globally by integrating or
  averaging all relevant quantities from pole to pole.  The exceptions
  are the strongest magnetised models.  Furthermore, a modification in
  which we compute the two timescales locally as a function of
  latitude accounts better for the deviations from spherical symmetry
  due to rapid rotation, in particular the flattening of the \nusp and
  the enhancement of the neutrino heating near the poles.  The
  agreement can be further improved by including the effects of
  lateral mixing between different angles, which limits the times a
  fluid element can be exposed to the most intense heating close to
  the poles.
\item The explosion criterion based on these two timescales
  corresponds to the condition that the total (MHD plus gravitational)
  energy of the gas in the gain layer be positive, \ie the gas be
  unbound, for an explosion to develop.  Our models, however, show
  sometimes large unbound regions long before the shock is revived.  A
  way to account for this delay would be to consider the Bernoulli
  parameter, \ie the specific total enthalpy, of the gas and compare
  it to the specific ram pressure ahead of the shock.  If a fluid
  element reaches a Bernoulli parameter exceeding the specific ram
  pressure while it falls through the gain layer an explosion is
  possible.  Since the former increases due to neutrino heating, the
  final explosion criterion is similar to that based on the advection
  and heating timescales.  However, by raising the threshold for shock
  revival, it agrees better with several simulations where the latter
  criterion is overly optimistic.
\item These two ways of analysing the explosion do not explicitly
  include magnetic fields.  Thus, their agreement with
  magnetorotational explosions is rather bad.  In our axisymmetric
  models, this kind of explosion develops out of a column of strong
  radial field located at the poles that connects the PNS to the
  immediate post-shock layers.  The shock runaway starts when the
  \Alfven speed exceeds the advection velocity.  Hence, we add an
  additional criterion by comparing the advection timescale to the
  timescale for the propagation of \Alfven waves through the gain
  layer.  Magnetorotational explosions start once the former exceeds
  the latter.  We note that the magnetic column suppresses angular
  motions flows, therefore the inclusion of lateral mixing as in the
  other two criteria is not required. 
\item We have verified that the most salient properties of a
    limited set of models also stand in 3D. We find that the 3D
    versions of models which possess the same rotational profile as
    the stellar evolution progenitors and either the same or larger
    magnetic fields, yield explosions with rather similar qualitative
    properties. A more quantitative analysis has been postponed to
    when we have higher resolution versions of the 3D models
    computed over longer post-bounce times.
\end{itemize}

We terminate by addressing the main limitations of the present study.
The most important drawback is certainly the restriction of
axisymmetry.  The amplification of magnetic fields, the dynamics of
the explosion, and the development of several instabilities can be
quite different in three-dimensional geometry.  Among the effects
depending on three-dimensional dynamics, we note in particular the
spiral modes of the SASI
\citep[\eg][]{Blondin_Shaw__2007__apj__Linear_Growth_of_Spiral_SASI_Modes_in_CCSNe,Fernandez__2010__apj__TheSpiralModesoftheStandingAccretionShockInstability,Hanke_et_al__2013__apj__SASIActivityinThree-dimensionalNeutrino-hydrodynamicsSimulationsofSupernovaCores,Guilet_Fernandez__2014__mnras__AngularmomentumredistributionbySASIspiralmodesandconsequencesforneutronstarspins},
MRI-driven turbulence and the possible dynamos
\citep{Moesta_et_al__2015__nat__Alarge-scaledynamoandmagnetoturbulenceinrapidlyrotatingcore-collapsesupernovae,Guilet_Mueller__2015__mnras__Numericalsimulationsofthemagnetorotationalinstabilityinprotoneutronstars-I.Influenceofbuoyancy,Masada_et_al__2015__apjl__MagnetohydrodynamicTurbulencePoweredbyMagnetorotationalInstabilityinNascentProtoneutronStars,Sawai_Yamada__2016__apj__TheEvolutionandImpactsofMagnetorotationalInstabilityinMagnetizedCore-collapseSupernovae,Guilet_et_al__2017__mnras__Magnetorotationalinstabilityinneutronstarmergers:impactofneutrinos},
and the low-$T/W$ instability of differentially rotating cores
\citep[\eg][]{Ott_et_al__2005__apjl__One-armed_Low-TW_Spiral_Instability,Kuroda_et_al__2014__prd__Gravitationalwavesignaturesfromlow-modespiralinstabilitiesinrapidlyrotatingsupernovacores,Takiwaki_et_al__2016__mnras__Three-dimensionalsimulationsofrapidlyrotatingcore-collapsesupernovae:findinganeutrino-poweredexplosionaidedbynon-axisymmetricflows}.
The first three-dimensional simulations run so far, albeit at reduced
grid resolution, show outflows that develop similarly to the
axisymmetric versions of the same models and thus seem to alleviate
the concerns, but some caution remains appropriate before drawing
overarching conclusions from the so far limited number of models.  We
will gradually increase the number of simulations, though the required
very long evolutionary times will remain a limiting factor.  In
comparison, our use of an approximate pseudo-GR potential instead of
full GR seems of minor importance.  Hence, our efforts for improving
upon this work should concentrate on simulating models in full
three-dimensional geometry, for which we are planning to address
selected issues in different stages of the evolution.

Finally, we stress that we have improved previously existing criteria
to understand the physical conditions at the onsed of the explosion
of high-mass stars. The combination of \Alfven and heating time
scales, suitably blured to account for the mixing properties in the
gain layer outperforms any of the previously existing explosion
criteria. It accomodates the very wide range of potential explosion
dynamics in high-mass stars (including rotation and magnetic fields),
specially, in CCSNe produced in highly compact cores.

\section{Acknowledgements}
\label{Sek:Ackno}

This work has been supported by the Spanish Ministry of Economy
Finance (AYA2015-66899-C2-1-P) and the Valencian Community
(PROMETEOII/2014-069).  MO acknowledges support from the European
Research Council under grant EUROPIUM-667912, and from the the
Deutsche Forschungsgemeinschaft (DFG, German Research Foundation) --
Projektnummer 279384907 -- SFB 1245.  We also thank the support from
the COST Actions PHAROS CA16214 and GWverse CA16104.  The computations
were performed under grants AECT-2016-1-0008, AECT-2016-2-0012,
AECT-2016-3-0005, AECT-2017-1-0013, AECT-2017-2-0006,
AECT-2017-3-0007, AECT-2018-1-0010, AECT-2018-2-0003,
AECT-2018-3-0010, and AECT-2019-1-0009 of the Spanish Supercomputing
Network on clusters \textit{Pirineus} of the Consorci de Serveis
Universitaris de Catalunya (CSUC), \textit{Picasso} of the Universidad
de M{\'a}laga, and \textit{MareNostrum} of the Barcelona
Supercomputing Centre, respectively, and on the clusters
\textit{Tirant} and \textit{Lluisvives} of the Servei d'Inform\`atica
of the University of Valencia.

\end{document}

%% file: explprop-table.tex
\begin{table*}
  \begin{center}
    \label{table1}
    \setlength\tabcolsep{3.5pt}
    \begin{filecontents*}{explprops-2.dat}

       Collapse-3noB    1.1900000000e+02    1.6744656886e+00    1.7040647555e+00    1.3985146517e+00    3.8729783392e+00    0.0000000000e+00    1.3711853370e+02    8.8734206961e+00    1.3458429401e+01    6.5932067046e-01    0.0000000000e+00    0.0000000000e+00
          s20-2    3.0500000000e+02    1.8588065579e+00    3.7159649133e-01    3.7818320993e-01    1.4151833606e-01    6.8966562479e+00    9.5765014230e+01    6.6139608231e+00    4.3749740340e+00    1.5117714463e+00    3.4154847645e+00    3.4299741877e+01
          s20-3    1.0800000000e+02    1.6501611784e+00    1.4165831657e+00    1.5609090485e+00    2.7840404500e+00    2.5234576633e+01    1.2162244797e+02    9.4697752732e+00    2.0958633281e+01    4.5183171757e-01    1.0151765398e+01    5.1335115193e+01
             35OC-RO    1.7800000000e+02    1.8577730247e+00    1.7612875056e+00    1.6929017533e+00    1.4899768308e+01    1.3525615866e+02    1.3913077768e+02    7.5578115898e+00    1.7645950174e+01    4.2830289756e-01    4.4985162749e+00    1.8818280838e+01
            35OC-RO2    6.0000000000e+01    1.5737223616e+00    2.8479701657e+00    2.6080533263e+00    7.2128037383e+00    6.4691359239e+01    1.0664550124e+02    1.0097042277e+01    8.7797062589e+01    1.1500432907e-01    2.3239817130e+00    1.6764489168e+01
            35OC-Rp2    1.1000000000e+02    1.6787340773e+00    1.9284715124e+00    1.9875044801e+00    7.2859038149e+00    3.9810387779e+01    1.2831208871e+02    2.3249689409e+01    6.3663158698e+01    3.6519848975e-01    5.8643538061e+00    1.6523979879e+01
            35OC-Rp3    8.0000000000e+01    1.5922194438e+00    1.7407821329e+00    2.1515519386e+00    5.6892155137e+00    3.9765307466e+01    1.0882261173e+02    1.9608911557e+01    1.1053906083e+02    1.7739350605e-01    4.8957815108e+00    2.1809321555e+01
            35OC-Rp4    5.0000000000e+01    1.5174456189e+00    3.5542072535e+00    3.5197293490e+00    4.4644707238e+00    3.7000320809e+01    9.9210721203e+01    1.6046182113e+01    1.1384758798e+02    1.4094441874e-01    5.1866681206e+00    1.5462275337e+01
            35OC-RRw    3.4300000000e+02    2.0994648607e+00    7.7359419473e-01    7.5763474862e-01    4.0188519505e+01    1.2490521170e-01    9.4243181103e+01    1.3593971942e+01    2.9863083122e+01    4.5520992882e-01    2.9873244586e-01    7.5257068695e+00
            35OC-Rs    2.0000000000e+01    1.2235819623e+00    4.3261079823e+00    7.9029629749e+00    3.0465233130e+00    6.9157745762e+01    9.1434777011e+01    4.0455650506e+00   6.4741143430e+01   6.2488316337e-02    1.3615713695e+01    1.9371409523e+01
             35OC-Rw    3.7800000000e+02    2.1465180949e+00    1.1338257351e+00    5.7783211488e-01    3.5562944048e+01    3.1617252230e+01    1.0062617125e+02    6.6788361945e+00    1.0269319088e+01    6.5036796864e-01    2.2750921536e+00    1.0174745027e+01
             35OC-Sw    4.1000000000e+02    2.1700148430e+00    3.3702592714e-01    4.1729881499e-01    5.5360402558e+00    3.5791429177e+00    1.1204053711e+02    5.3548188467e+00    2.7844491531e+00    1.9231160464e+00    2.2601055449e+00    1.7243797040e+01
             35OB-RO    4.2500000000e+02    2.1280513424e+00    1.0772410470e+00    1.1363076076e+00    2.0840082276e+01    7.0498150543e+01    1.6803409100e+02    6.0866648182e+00    1.0115443306e+01    6.0172002691e-01    4.6447976624e+00    1.0723844730e+01
            35OB-RRw    2.2380000000e+03    3.3526147324e+00    6.0962240932e-01    5.8885811086e-01    1.6866548865e+02    3.0865684122e+02    7.9884144603e+01    1.1148993134e+01    2.1286700931e+01    5.2375392366e-01    9.2533073782e+00    2.5111505413e+01
              z35-Rw    1.2210000000e+03    2.7339447334e+00    2.1148151249e-01    4.7479478131e-01    8.3406383235e+01    1.4890700295e-01    5.8829754775e+01    1.4914521053e+01    6.0871241228e+00    2.4501752801e+00    2.2069521927e-01    1.3640047173e+00
              z35-Sw    8.2200000000e+02    2.5644676359e+00    4.8326161253e-01    5.6719579890e-01    3.5103129814e+01    1.4534440043e-01    1.2377651140e+02    7.8589528105e+00    3.1988975615e-01    2.4567691398e+01    1.6221880757e-01    1.5228858701e+00
\end{filecontents*}
\pgfplotstabletypeset[
      multicolumn names, 
      col sep=space, 
      string type,
      display columns/0/.style={
		column name=Model, 
		column type=l,
              },
      display columns/1/.style={
         sci zerofill, sci sep align, sci 10e,
                precision=3,
                column name = $t_{\mathrm{exp}}$
              },
      display columns/2/.style={
        sci zerofill, sci sep align, 
                precision=2,
                column name = $M_{\pnss}^{\EE}$ 
              },
      display columns/3/.style={
        sci zerofill, sci sep align, 
                column name = $\dot{M}_{\pnss}^{\EE}$ 
              },
      display columns/4/.style={
        fixed,sci sep align, 
                column name = $F^{\textsc{m}}_{\mathrm{gain}}$ 
              },
      display columns/5/.style={
        fixed,sci sep align, 
                column name = $\Erot^{\pnss;\EE}$
              },
      display columns/6/.style={
                column name = $\Emag^{\pnss;\EE}$,
        fixed,sci sep align, 
              },
      display columns/7/.style={
        fixed,sci sep align, 
        precision=0,
                column name = $L^{\EE}_{\nu_e + \bar{\nu}_e}$
              },
      display columns/8/.style={
        fixed,sci sep align, 
                column name = $\tau_{\mathrm{adv}}^{\EE}$
              },
      display columns/9/.style={
        fixed,sci sep align, 
                column name = $\tau_{\mathrm{heat}}^{\EE}$
              },
      display columns/10/.style={
        fixed,sci sep align, 
                column name = $\displaystyle\frac{\tau_{\mathrm{adv}}^{\EE}}{\tau_{\mathrm{heat}}^{\EE}}$
              },
      display columns/11/.style={
        fixed,sci sep align, 
                column name = $\langle c^{\EE}_{\textsc{a}} \rangle$
              },
      display columns/12/.style={
        fixed,sci sep align, 
                column name = $c^{\mathrm{pole;}\EE}_{\textsc{a}}$
              },
      every head row/.style={
        before row={\toprule}, 
        after row={
          &
          $[\mathrm{ms}]$ & &
          $[M_{\mathrm{\odot}}]$ & &
          $[\frac{M_{\mathrm{\odot}}}{\mathrm{s}}]$ & &
          $[\frac{M_{\mathrm{\odot}}}{\mathrm{s}}]$ & &
          $[\mathrm{foe}]$ & &
          $[0.01 \, \mathrm{foe}]$ & &
          $[\frac{\mathrm{foe}}{\mathrm{s}}]$ & &
          $[\mathrm{ms}]$ & &
          $[\mathrm{ms}]$ & &
          & &
          $[10^8 \frac{\mathrm{cm}}{\mathrm{s}}]$ & &
          $[10^8 \frac{\mathrm{cm}}{\mathrm{s}}]$ & 
          \\ 
          \midrule} 
      },
      every last row/.style={after row=\bottomrule}, 
      ]{explprops-2.dat} 
  \end{center}
  \caption{
    Properties of the models at the time of the onset of the explosion.
    From left to right, the columns display
    the model name,
    the time of shock revival,
    the mass of the PNS and the rate at which it grows,
    its rotational and magnetic energies,
    the combined luminosities of the electron-type neutrinos,
    the advection and heating timescales and their ratio,
    the volume average of the  \Alfven speed in the gain layer and
    its mean value on the two poles.
    Models that fail to explode are excluded.
  }
  \label{Tab:Exprops}
\end{table*}